\documentclass[12pt]{iopart}

\expandafter\let\csname equation*\endcsname\relax
\expandafter\let\csname endequation*\endcsname\relax
\usepackage[vcentermath]{youngtab}
\usepackage{amsfonts}
\usepackage{subfig}
\usepackage{graphicx}
\usepackage{amsmath}
\usepackage{amssymb}
\usepackage[square, numbers, sort&compress, nonamebreak]{natbib}
\usepackage{color}\usepackage{url}
\usepackage[bookmarks, breaklinks]{hyperref}

\usepackage{xcolor} % allow for colour comments in the text
\newcommand{\red}[1]{{\color{red} #1}}
\newcommand{\blue}[1]{{\color{blue} #1}}

\newcommand{\newblock}{}

\newcommand{\vect}[1]{\mathbf{#1}}%{\bm{#1}}

\definecolor{johannesstyle}{rgb}{0.0,0.8,1.0}

\usepackage[normalem]{ulem}

\begin{document}

\title[Strongly correlated one-dimensional Bose-Fermi quantum mixtures]{Strongly correlated one-dimensional Bose-Fermi quantum mixtures: symmetry and correlations}

\author{Jean Decamp}
\address{Universit\'e C\^ote d'Azur, CNRS, Institut de Physique de Nice, France}

\author{Johannes J\"unemann}
\address{Johannes Gutenberg-Universit\"at, Institut f\"ur Physik, Staudingerweg 7, 55128 Mainz, Germany}
\address{Graduate School Materials Science in Mainz, Staudingerweg 9, 55128 Mainz, Germany}

\author{Mathias Albert}
\address{Universit\'e C\^ote d'Azur, CNRS, Institut de Physique de Nice, France}

\author{Matteo Rizzi}
\address{Johannes Gutenberg-Universit\"at, Institut f\"ur Physik, Staudingerweg 7, 55128 Mainz, Germany}

\author{Anna Minguzzi}
\address{Universit\'e Grenoble-Alpes, CNRS, LPMMC, BP166, F-38000 Grenoble, France}

\author{Patrizia Vignolo}
\address{Universit\'e C\^ote d'Azur, CNRS, Institut de Physique de Nice, France}

\date{\today}

\begin{abstract}
We consider multi-component quantum mixtures (bosonic, fermionic, or mixed) with strongly repulsive contact interactions in a one-dimensional harmonic trap. In the limit of infinitely strong repulsion and zero temperature,  using the class-sum method, we study the  symmetries of the spatial wave function of the mixture. We find that the ground state of the system has the most symmetric spatial wave function allowed by the type of mixture. This provides an example of the generalized Lieb-Mattis theorem. Furthermore, we show that the symmetry properties of the mixture are embedded in the large-momentum tails of the momentum distribution, which we evaluate both at infinite repulsion by an exact solution and at finite interactions using a numerical DMRG approach. This implies that an experimental measurement of the Tan's contact would allow to unambiguously determine the symmetry of any kind of multi-component mixture. 
\end{abstract}

\pacs{05.30.-d,67.85.-d,67.85.Pq}

%\maketitle

%%%%%%%%%%%%%%%%%%%%%%%%%%%%%%%%%%%%%%%%%%%%%%%%%%%%%%%%%%%%%%%%%%%%%%

\section{Introduction}

Ultracold atom experiments allow to engineer and probe, with an incredible and always-improving precision, a yet inaccessible variety of many-body quantum systems \cite{Bloch2008}. One-dimensional (1D) models, which display several unique features associated to the reduced dimensionality \cite{Giamarchi_book}, are the object of intense  theoretical and experimental interest. Quantum gases in one dimension can be realized in actual experiments by trapping atoms in tight optical waveguides \cite{Moritz2003,Paredes2004,Clement2009,Pagano2014}. Moreover, these experiments offer the possibility to tune the interactions between the atoms \cite{Bloch2008,Olsh98}, and hence to  access the strongly correlated regime for various kinds of quantum mixtures. For instance, a strongly repulsive two-component 1D Fermi gas was realized in \cite{Zurn2012}, and a mixture with up to six fermionic components was realized using Ytterbium atoms \cite{Pagano2014}. Other experiments with Ytterbium isotopes offer promising perspectives for the realization of strongly interacting 1D Bose-Fermi mixtures \cite{Fukuhura09}.

In multi-component Bose-Fermi quantum mixtures, the exchange symmetry between identical particles is fixed by their bosonic or fermionic nature, but not between distinguishable particles. Thus, a natural question is to find how to characterize, both theoretically and experimentally, the global exchange symmetry of the many-body wave-function. This fundamental property is directly related to the magnetic properties of the system \cite{Deuretzbacher2014,Volosniev2014,Sutherland68}. Therefore, strongly interacting 1D atomic mixtures appear to be a perfect experimental and theoretical playground for studying quantum magnetism \cite{Cazalilla2014}. They also offer the opportunity to study itinerant magnetism phases, i.e.~magnetism without a lattice \cite{Massignan2015}.

A typical feature in  strongly interacting gases is the so-called fermionization: at increasing repulsive interactions, particles which are not subjected to the Pauli principle tend to avoid each other. In the limit of infinite repulsion, contact interactions mimic the Pauli exclusion principle, e.g.\@ they induce zeros in the many-body wave function when two particles meet. This phenomenon allows  the construction of an exact  solution for any type of quantum mixture through a mapping onto a  non-interacting spinless Fermi gas \cite{Gir1960,Deuretzbacher,GirMin2007,Fang2011,Volosniev2014,Zinner2017}. 

In this work, we focus on 1D multi-component Bose-Fermi quantum mixtures with strongly repulsive contact interactions, taking the experimentally relevant case of a harmonic external confinement. In the limit of infinite interactions, we provide an exact solution of the multi-component mixture with arbitrary number of components using the method developed in \cite{Volosniev2014,Zinner2017}. In order to address the symmetry of the many-body wave function of the ground and excited states we use the class-sum method. We show  that the mixtures obey a generalized version of the Lieb-Mattis theorem \cite{LiebMattisPR}. Namely, as obtained for multi-component fermions in \cite{Decamp2016}, the spatial ground state wave function of each system is the  most symmetric one. Our analysis of the symmetries also  confirms the ordering of the excited states recently obtained in \cite{Pan2017} for fermionic mixtures and extends this concept to the case of Bose-Fermi mixtures. Furthermore, we explore how symmetry affects the properties of the system in coordinate and momentum space by studying density profiles and momentum distributions. For this purpose we combine the exact solution at infinite interactions with numerical Density Matrix Renormalization Group (DMRG) calculations at finite interactions. This allow us to analyze the symmetry structure of the mixture in the fermionization process. In particular, we show that a precise measurement of the asymptotic momentum distribution would allow to probe the symmetry of the mixture, thus generalizing our previous result on multi-component fermions \cite{Decamp2016-2}. 

The paper is organized as follows: In Section \ref{sec:Model}, we give a description of our model and of its solution. Then, in Section \ref{sec:Symmetry}, we provide a detailed  description of the class-sum method that we use to extract the exchange symmetry of the many-body wave function.  In Section \ref{sec:Correlations}, we provide our results for the density profiles and momentum distributions. Finally, Sec. \ref{sec:Concl} gives our concluding remarks and perspectives.

%%%%%%%%%%%%%%%%%%%%%%%%%%%%%%%%%%%%%%%%%%%%%%%%%%%%%%%%%

\section{Model and general solution}
\label{sec:Model}

\medskip

We consider a mixture of $N_B=N_1^B+\ldots+N_b^{B}$ bosons and $N_F=N_1^{F}+\ldots+N_f^{F}$ fermions,  with   $N=N_B+N_F$,   divided in $b$ and $f$ flavors (\textit{e.g.} spin components), all having the same mass $m$. For a Bose-Fermi mixture this is always an approximation, which is appropriate e.g.\@ for mixtures of isotopes. This type of mixtures is experimentally accessible with Ytterbium atoms \cite{Fukuhura09} for instance. The particles are confined  in a tight atomic waveguide so that their motion can be considered one-dimensional.   Their positions are given by the coordinates $x_1^{B,1},\ldots,x_{N_B}^{B,b},x_{N_B+1}^{F,1},\dots,x_{N}^{F,f}$. For the sake of simplicity of notations, the exponents specifying the type and flavor of the particle will be from now on omitted. All the particles are also subjected to the same one-dimensional confinement   $V(x)=\frac{1}{2} m \omega^2 x^2$, which could be realized e.g.\@ in cigar-shaped optical traps. The interactions between the particles are modelled via the pseudo-potential
 $V_{\mathrm{int}}(x,y)=g\delta(x-y)$. Here $g=-2\hbar^2/m a_{1D}$, and $a_{1D}$ is the 1D effective scattering length \cite{Olsh98}, which accurately describes the $s$-wave scattering dominated collisions in a restricted geometry at low enough energies and densities~\cite{Bloch2008}.
We assume that the boson-boson, boson-fermion and fermion-fermion interactions are characterized by the same interaction strength $g$. In the fermionic case, no $s$-wave collisions occur among fermions belonging to the same flavor. The Hamiltonian describing the system is then given by
\begin{equation} \label{eq:ham}
  H =  \sum_{i=1}^{N} \left( - \frac{\hbar^2}{2m} \frac{\partial^2}{\partial x_i^2} + \frac{1}{2} m 
  \omega^2 x_i^2 \right) \
  + g\sum_{1\le i<j\le N}\delta(x_i-x_j).
\end{equation}

The interaction term in $H$ is equivalent to imposing the cusp condition on the many-body wave function $\Psi=\Psi(x_1, \ldots, x_N)$,
\begin{equation}
\label{eq:cusp}
\partial_x \Psi(x=0^+) - \partial_x \Psi(x=0^-) = \frac{m g}{\hbar^2} \, \Psi(x=0) \ ,
 \end{equation}
where $x=x_i-x_j$. 

\subsection{Exact many-body wave function at infinite repulsion}
As a consequence of the cusp condition, in the strongly correlated limit $g\to\infty$, the many-body wave function  vanishes whenever $x_i=x_j$, leading to fermionization features. In this limit one can build an exact many-body wave function satisfying the cusp condition by taking  \cite{Volosniev2014,Deuretzbacher}
\begin{equation}
\label{eq:Psi}
\Psi(x_1,\ldots,x_N)=\sum_{P\in S_N}a_P\theta_P(x_1,\ldots,x_N)\Psi_A(x_1,\ldots,x_N),
\end{equation}
where $S_N$ is the permutation group of $N$ elements, $\theta_P(x_1,\ldots,x_N)$ is equal to $1$ if $x_{P(1)}<\cdots<x_{P(N)}$  (coordinate sector, later indicated as $(P(1), P(2), ...P(N))$) and $0$ otherwise, and $\Psi_A$ is the fully antisymmetric fermionic  wave function  $\Psi_A(x_1,\ldots,x_n)=\frac{1}{\sqrt{N!}}\mathrm{det}[\phi_{i-1}(x_j)]_{i,j=1,\ldots,N}$  where $\phi_0,\ldots,\phi_{N-1}$ are the eigenfunctions 
of the single particle Hamiltonian $H_1=-(\hbar^2/2m)\partial_x^2+V(x)$.
The coefficients $a_P$ in Eq.(\ref{eq:Psi}) corresponding to exchange of particles belonging to the same bosonic (fermionic) component of the mixture  are such that $a_P=1$ ($a_P=-1$) respectively.
Hence, the number of independent coefficients
is reduced to $D_{N,\mathrm{b+f}}=\frac{N!}{N_1!\ldots N_{\mathrm{b+f}}!}$. This allows to extremely reduce the cost of calculations, by identifying sectors that are equal
\textit{modulo} permutations of identical particles. This smaller basis is often called the snippet basis \cite{Deuretzbacher,Fang2011}. In this
paper, we will switch from one formalism to the other depending on whether a distinction between identical bosons is necessary or not.

In order to find the coefficients $a_P$ we use the variational approach developed in \cite{Volosniev2014}.
This is based on a strong-coupling expansion of the energy to order  $1/g$, i.e.\@ by setting   $E=E_A-K/g$ where $E_A$ is the fermionic energy associated with $\Psi_A$. Using the Hellmann-Feynman theorem together with (\ref{eq:cusp}) we obtain
\begin{equation} 
\label{eq:K}
K=g\left<H_{\mathrm{int}}\right>=\sum_{P,Q\in S_N}(a_P-a_Q)^2\alpha_{P,Q},
\end{equation}
where $\alpha_{P,Q}=\int \mathrm{d}x_1\ldots\mathrm{d}x_N \theta_{\mathrm{Id}}(x_1,\ldots,x_N) \delta(x_k~-~x_{k+1}) 
\left[\partial\Psi_A/\partial x_k\right]^2\equiv\alpha_k$
if $P$ and $Q$ are equal up to a transposition $\tau_k$ of two consecutive particles that are either distinguishable particles or indistinguishable bosons
at positions $k$ and $k+1$, and $0$ otherwise (see \cite{Decamp2016,Deuretzbacher2014}
for computational methods). Note that the spatial parity of the trap implies that $\alpha_k=\alpha_{N-k}$ for all $k$. Then, the variational condition 
$\partial K/\partial a_P=0$ is shown to be equivalent to a simple diagonalization problem
$V\vect{a}_{\mathrm{snip}}=K\vect{a}_{\mathrm{snip}}$, where $\vect{a}_{\mathrm{snip}}$ is the vector of the $D_{N,\mathrm{b+f}}$ independent $a_i$ coefficients and $V$ is a 
$D_{N,\mathrm{b+f}}\times D_{N,\mathrm{b+f}}$ matrix defined 
in the snippet basis by
\begin{equation}
\label{volosniev}
V_{ij}=\left\{
\begin{array}{ll}
 -\alpha_{i,j} & \mbox{if } i\ne j \\ \sum_{\mathrm{d},k\ne i}\alpha_{i,k}+2\sum_{\mathrm{b},k\ne i}\alpha_{i,k}& \mbox{if  } i=j
\end{array}
\right.,
\end{equation}
where the index $\mathrm{d}$ means that the sum has to be taken over snippets $k$ that transpose distinguishable particles as compared to snippet $i$, 
while $\mathrm{b}$ means that the sum is taken over sectors that transpose identical bosons. The highest eigenvalue of $V$  yields the ground state of the system, whereas the other eigenvalues correspond to all the excited states that belong to the same $D_{N,\mathrm{b+f}}$-degenerate manifold at $g\rightarrow\infty$.

\section{Symmetry of the mixture}
\label{sec:Symmetry}

In this section, we study the spatial symmetry of our solution under exchange of particles. To this purpose we use the class-sum method,  inspired from nuclear physics \cite{Talmi,Novolesky1995}, and previously used in \cite{Fang2011,Decamp2016,Decamp2016-2}.

\subsection{General method}

We consider a $N$-particle wave function $\Psi(x_1,\dots,x_N)$, identified   by a vector $\vect{a}_{\mathrm{sect}}=(a_1,\dots,a_{N!})$ as in Eq.~(\ref{eq:Psi}).  
We want to determine its permutational symmetry, i.e. find to which irreducible 
representation(s) of the permutation group $S_N$ it belongs. Each irreducible representation can be labeled by a Young diagram, where, according to the standard notation,
 boxes in the same column (resp. row) correspond to an antisymmetric 
(resp. symmetric) exchange of particles \cite{Hamermesh_book}. Examples of Young diagrams and their corresponding symmetries for $N=4$ are given 
in Table \ref{tab:young}.
Using the Young diagrams, it is possible to know \textit{a priori}, given a configuration $N_1^B+\ldots+N_b^{B}+N_1^{F}+\ldots+N_f^{F}$, which irreducible
representations are possible for $\Psi$, compatible with the antisymmetrization of fermions belonging to the same component,
and symmetrization for same-component bosons. More precisely, in order to build the standard Young tableaux for a given mixture, we index by a letter the particles of each given  species. We therefore order the species by decreasing order and label them alphabetically. We then label the boxes of the Young diagrams imposing that the entries of each row and column are always increasing (in weak sense), 
and with the symmetry constraint that no more than one same-component bosonic/fermionic label can be present in one column/row.
Examples are provided for $4$-particle mixtures in Table~\ref{tab:young}.
\begin{table}
%\subfloat[Young diagrams in terms of exchange symmetry]
{
\begin{tabular}{l|r}
Exchange symmetry & Young diagram(s) \\\hline
symmetric & {\tiny\yng(4)}  \\
antisymmetric & {\tiny\yng(1,1,1,1)} \\
mixed & {\tiny\yng(3,1)} {\tiny\yng(2,2) {\tiny\yng(2,1,1)}}
\end{tabular}
}
%\subfloat[Standard Young tableaux at fixed mixture]
{
\begin{tabular}{l|r}
Mixture & Young tableaux \\\hline
$2^B(a)+2^B(b)$ & {\tiny\young(aabb)} {\tiny\young(aab,b)} {\tiny\young(aa,bb)}  \\
$2^F(a)+2^F(b)$ & {\tiny\young(a,a,b,b)} {\tiny\young(ab,a,b)} {\tiny\young(ab,ab)} \\
$2^B(a)+2^F(b)$ & {\tiny\young(aab,b)} {\tiny\young(aa,b,b)}
\end{tabular}
}
\caption{Young diagrams and standard Young tableaux for different two-component mixtures with $N=4$. The particles of the species are labeled $a$ resp. $b$, and can be either bosonic or fermionic.}
\label{tab:young}
\end{table}
We also recall that the number of standard Young tableaux corresponding to a given Young  diagram is equal to the dimension
of the associated irreducible representation \cite{Kerber}.
The number $d_{\lambda}$ 
of possible standard Young tableaux associated with a Young diagram $\lambda$ is
given by the Hook-length formula
\begin{equation}
d_{\lambda}=\prod_{(i,j)\in\mu}\frac{N!}{h_{\lambda}(i,j)},
\end{equation}
where $h_{\lambda}(i,j)$ is equal to the number of cells below the box  $(i,j)$ + the number of cells at the right of the box $(i,j)$ + 1. Interestingly, the  dimension $d_{\lambda}$  counts the number of energy eigenstates having the  symmetry $\lambda$  \cite{Harshman2014,Harshman2016}.

In order to obtain the symmetry associated with a given wave function $\Psi(x_1,\dots,x_N)$ belonging to the degenerate manifold, we define a set of $N!\times N!$ matrices whose eigenvalues are directly
connected to the irreducible representations of $S_N$, namely the conjugacy class-sums \cite{Kerber,Liebeck}. Considering cycles $\sigma$ of length
$|\sigma|=p$, which  permute $p$ elements of $\{1,\dots,N\}$ in a cyclic way, the $p$-cycle class-sum $\Gamma^{(p)}$ is represented  in the  coordinate-sector basis of $N$ non-interacting fermions as
\begin{equation} \label{eq:class}
\Gamma^{(p)}=\sum_{\sigma, |\sigma|=p}M_{\sigma}.
\end{equation}
$M_{\sigma}$ is an $N!\times N!$ matrix whose elements are $\left(M_{\sigma}\right)_{PQ}=(-1)^{|\sigma|-1}\delta_{P,\sigma\circ Q}$, and 
the factor $(-1)^{|\sigma|-1}$  is due to the inherent antisymmetry of the basis. 
As an example, for $N=3$, taking as  basis $\left((1,2,3),(1,3,2),(2,1,3),(2,3,1),(3,1,2),(3,2,1)\right)$, one has  
\begin{equation}
\Gamma^{(2)}=\begin{pmatrix} 0 & -1 & -1 & 0 & 0 & -1 \\
-1 & 0 & 0 & -1 & -1 & 0 \\ 
-1 & 0 & 0 & -1 & -1 & 0 \\
 0 & -1 & -1 & 0 & 0 & -1 \\
  0 & -1 & -1 & 0 & 0 & -1 \\
-1 & 0 & 0 & -1 & -1 & 0 \\
\end{pmatrix},
\end{equation}
and
\begin{equation}
\Gamma^{(3)}=\begin{pmatrix} 0 & 0 & 0 & 1 & 1 & 0 \\
0 & 0 & 1 & 0 & 0 & 1 \\ 
0 & 1 & 0 & 0 & 0 & 1 \\
1 & 0 & 0 & 0 & 1 & 0 \\
1 & 0 & 0 & 1 & 0 & 0 \\
0 & 1 & 1 & 0 & 0 & 0 \\
\end{pmatrix}.
\end{equation}

The key point of our method is that the class-sum eigenvalues $\gamma^{(p)}$ are directly related to the irreducible representations of $S_N$: 
given a Young diagram $\lambda=[\lambda_1,\ldots,\lambda_n]$, where $\lambda_i$ is the number of boxes in row $i$, the following relation holds \cite{Katriel1993}:
\begin{equation} \label{eq:classev}
\gamma^{(p)}=\frac{1}{p}\sum_{i=1}^n\mu_i(\mu_i-1)\ldots(\mu_i-p+1)\prod_{j\ne i}\frac{\mu_i-\mu_j-p}{\mu_i-\mu_j}.
\end{equation}
where $\mu_i=\lambda_i-i+n$. \footnote{Note that there is a notation misprint in Ref.~\cite{Katriel1993} where $n$ refers to the total number of atoms instead of the total number of rows.}
The fundamental reason behind the above connection (\ref{eq:classev}) between class sums and Young tableaux
is that there is a one-to-one correspondence of the eigenvalues of a class-sum  and the  characters of the irreducible representations of $S_N$ \cite{Novolesky1995,Liebeck}.
Once one considers the relation between the irreducible representations of $S_N$ and the ones of $SU(\kappa)$ (for the case of $\kappa$-species mixtures) \cite{Hamermesh_book}, it is possible to connect the eigenvalues of the class-sum operators and those of the $SU(\kappa)$ Casimir operators. The latter are the ones that, roughly speaking, describe the character of $SU(\kappa)$ irreps. and thus generalise the $SU(2)$ concept of spin length.

Using the above properties, in order to determine the symmetry of the wave function  $\Psi$ of a given mixture we proceed as follows: first, we determine the matrix  $\Gamma^{(2)}$ on the coordinate-sector basis and compute its eigenvalues $\gamma^{(2)}$. Then,  we list all the Young diagrams allowed within the given type of mixture, and   using Eq.~(\ref{eq:classev}), we associate each eigenvalue  $\gamma^{(2)}$ to a given Young diagram. If different diagrams correspond to the same $\gamma^{(2)}$, we  repeat this operation to higher order in the size $p$ of the cycles until  there are no longer degeneracies.  Finally, we expand the  wave function $\Psi$, represented in the same basis as the vector  $\vect{a}_{\mathrm{sect}}=(a_1,\dots,a_{N!})$, on the basis of the eigenvectors of   $\Gamma^{(p)}$ in order to determine the weights of the different irreducible representations, and hence the symmetry content of the wave function.

Finally, we would like to comment that the class-sum method has been here adapted  to the ansatz (\ref{eq:Psi}),
but it can be used as well for  other ansatz, e.g.\@  the Bethe ansatz  \cite{Yang67,Sutherland68}, as long as the $M_{\sigma}$ matrices in Eq.~(\ref{eq:class}) are defined correspondingly.

\subsection{Results and discussion}
The case of fermionic mixtures has already been studied in \cite{Decamp2016,Decamp2016-2}, and the case of simple Bose-Fermi mixtures $N_1^B+N_1^F$ with
$N_1^B=N_1^F$ was studied in \cite{Fang2011}. In this work, we extend these results to the more complex and general case of Bose-Fermi mixtures with more than one  bosonic/fermionic components. In particular, we study the cases $2^B+2^F$, $3^B+3^F$, $2^B+2^B+2^F$ and $2^B+2^F+2^F$. Note that we only need $p_{\mathrm{max}}=3$
in order to discriminate between {\tiny\yng(4,1,1)} and {\tiny\yng(3,3)} or between {\tiny\yng(3,1,1,1)} and {\tiny\yng(2,2,2)}, which are degenerate in $\Gamma^{(2)}$.
We focus on the ground states of a given symmetry configuration, in the sense that, given a symmetry $S$, they have the highest eigenvalue of $V$ denoted $K(S)$,  corresponding to the 
the lowest energy $E(S)$. Our results are
shown in Table \ref{tab:sym}.
\begin{table}
\begin{center}
\subfloat[Two-component mixtures.]{
\begin{tabular}{l|c|c|c}
Mixture & Symmetry & $\gamma^{(2)}$ & $K(S)/(\hbar^2\omega^2a_{ho})$ \\\hline
$2^B+2^F$ & {\tiny\yng(3,1)} & 2 & 10.66 \\
          & {\tiny\yng(2,1,1)} & -2 &  7.08 \\\hline
$3^B+3^F$ & {\tiny\yng(4,1,1)} & 3 & 30.37 \\
          & {\tiny\yng(3,1,1,1)} & -3 & 24.43 \\
\end{tabular}
}
\\
% \quad
\subfloat[Three-component mixtures: Young tableaux in black are the possible ones for both mixtures whereas the blue (red) ones are restricted to the $2^B+2^B+2^F$ ($2^B+2^F+2^F$) mixture.]{
\begin{tabular}{l|c|c|c}
Mixtures & Symmetry & $\left(\gamma^{(2)},\gamma^{(3)}\right)$ & $K(S)/(\hbar^2\omega^2a_{ho})$ \\\hline
\blue{$2^B+2^B+2^F$} & \blue{{\tiny\yng(5,1)}} & $\left(9,16\right)$ & 33.35 \\
\red{$2^B+2^F+2^F$} & {\tiny\yng(4,2)} & $\left(5,0\right)$ & 32.16 \\
          & \blue{{\tiny\yng(3,3)}} & $\left(3,-8\right)$ & 30.63 \\
          & {\tiny\yng(4,1,1)} & $\left(3,4\right)$ & 30.37 \\
          & {\tiny\yng(3,2,1)} & $\left(0,-5\right)$ & 28.96 \\
          & \red{{\tiny\yng(2,2,2)}} & $\left(-3,-8\right)$ & 24.97 \\
          & {\tiny\yng(3,1,1,1)} & $\left(-3,4\right)$ & 24.43 \\
          & {\tiny\yng(2,2,1,1)} & $\left(-5,0\right)$ & 22.69 \\
          & \red{{\tiny\yng(2,1,1,1,1)}} & $\left(-9,16\right)$ & 14.60 \\
\end{tabular}
}
\end{center}
\caption{\label{tab:sym}(Color online) Symmetries diagrams obtained for different states and different mixtures, with their corresponding class-sum eigenvalues $\left(\gamma^{(2)},\gamma^{(3)},\ldots,\gamma^{(p_{\mathrm{max}})}\right)$ and their energy slopes $K(S)$ in units of $\hbar^2\omega^2a_{ho}$.}
\end{table}
First, we observe that, as in the case of Fermi mixtures \cite{Decamp2016}, these states have a pure symmetry, \textit{i.e.} they belong to a unique irreducible representation. Moreover, a direct comparison of Table \ref{tab:sym}(a) and Table \ref{tab:sym}(b) shows that Bose-Fermi mixtures with more than two components have much complex 
symmetric structures than the two-component ones, the latter having always only two possible symmetries \cite{Fang2011}.
Another interesting observation is that a given symmetry $S$ corresponds to the same ground state energy slope $K(S)$, independently of the mixture. 
In that sense, the symmetry of the mixture appears to be a deeper physical characterization  than the actual choice of the mixture. 

Furthermore, it is clear in Table \ref{tab:sym} that diagrams with higher $K(S)$ or, equivalently, lower $E(S)$, are more ``horizontal'', or more ``symmetric''.
In the context of a two-spin component fermionic mixture, this observation is known as the Lieb-Mattis theorem and has very strong implications in the theory of ferromagnetism \cite{LiebMattisPR}. It states that the mixture with the lowest total spin, and therefore the most symmetric spatial wave-function, will display the lowest ground-state energy, which implies that the ground state is unmagnetized.
On the other hand, it was proven in \cite{Lieb2002} that the ground state of an interacting bosonic mixture is always fully polarized, which again is equivalent to the property that the spatial wave function is the most symmetric.
Our results show that the ground
state corresponds to the most symmetric spatial configuration allowed by the Bose-Fermi mixture, generalizing the results obtained in the case of purely 
fermionic mixtures in \cite{Decamp2016,Decamp2016-2},
in agreement with the second Lieb-Mattis Theorem (LMT II) \cite{LiebMattisPR}. 

The notion that a state $S$ is ``more symmetric'' than $S'$ is usually understood in the sense of the so-called
pouring principle, which in terms of Young diagrams  means that the diagram corresponding to $S$ can be poured into the diagram corresponding 
to $S'$ by moving boxes up and right \cite{LiebMattisPR}: e.g.\@ {\tiny\yng(4,2)} can be poured into {\tiny\yng(5,1)}.
The LMT II claims that $E(S)>E(S')$ if $S$ can be poured in $S'$.
Nevertheless the pouring principle has
its limitations: clearly {\tiny\yng(4,1,1)} and {\tiny\yng(3,3)} can not be compared according to this principle. Here, we show that these symmetries display
different interaction energies and are therefore comparable (\textit{e.g.} $K\left({\tiny\yng(4,1,1)}\right)<K\left({\tiny\yng(3,3)}\right)$), confirming the ordering 
 observed in \cite{Pan2017} in the case of a Fermi gas and extending the analysis to the case of Bose-Fermi mixtures. 

%%%%%%%%%%%%%%%%%%%%%%%%%%%%%%%%%%%%%%%%%%%%%%%%%%%%%%%%%

\section{Density profiles and momentum distributions}
\label{sec:Correlations}

This section is dedicated to the study of the effect of symmetries on the one-body density matrix $\rho_{\nu}(x,y)$ in the strongly interacting regime.  For any component $\nu\in{1,\ldots,\mathrm{b+f}}$, $\rho_{\nu}(x,y)$ is defined  by
\begin{equation}
\rho_{\nu}(x,y)=N_{\nu}\int\mathrm{d}x_2\ldots\mathrm{d}x_N\Psi^*(x,x_2,\ldots,x_N)\Psi(y,x_2,\ldots,x_N),
\label{rhonu}
\end{equation}
where the first coordinate of $\Psi$ is chosen to belong to species $\nu$. The one-body density matrix measures the first-order spatial coherence among  two particles of species $\nu$ in positions $x$ and $y$ respectively. From the knowledge of the one-body density matrix one can obtain  the density profile, according to 
\begin{equation}
n_{\nu}(x)=\rho_{\nu}(x,x).
\end{equation}
This  quantity measures 
the probability of finding a particle of species $\nu$ in position $x$. It is accessible in cold-atom experiments via \textit{in situ} 
imaging \cite{Bloch2008}.

The momentum distribution $n_\nu(k)$ of species $\nu$ is also directly extracted from the one-density matrix by Fourier transform, according to
\begin{equation}
\label{eq:defnk}
n_{\nu}(k) =  \frac{1}{2 \pi}\iint \mathrm{d}x\,\mathrm{d}y\,  \rho_{\nu}(x,y)e^{-ik(x-y)}.
\end{equation}
The momentum distribution is one of the most common observables in cold-atom experiments. All the components of a mixture can be separately measured  using spin-resolved time-of-flight techniques  \cite{Bloch2008,Lewenstein_book}.
The large-momentum tails of the momentum  distribution for each component of a mixture  follow a universal $k^{-4}$ law, as first proven for the bosonic Tonks-Girardeau  gas in 
\cite{Minguzzi02,Olshanii03}. 
The weights $\mathcal{C}_{\nu}=\lim_{k\to\infty}k^4n_{\nu}(k)$, known as Tan's contacts, are of primary importance because of their relations 
to many other physical quantities, as the two-point correlators or the interaction energy, in various quantum mixtures and trapping potential 
\cite{Tan2008a,Tan2008b, Tan2008c,Zwe11,Decamp2016-2}. One of these relations allows to link $\mathcal{C}_{\nu}$ to the interaction energy, and more specifically
to the part of the interaction energy to which species $\nu$ contributes.
More precisely, if we denote $H_{\mathrm{int},\nu\nu^\prime}$ the interaction energy between species $\nu$ and $\nu^\prime$, the following relation holds
\begin{equation}\label{eq:tanrel}
\mathcal{C}_{\nu}=\frac{m^2 g}{\pi\hbar^4} \sum_{\nu^\prime} (1+\delta_{\nu\nu^\prime}) H_{\mathrm{int},\nu\nu^\prime}.
\end{equation}

\subsection{Exact results at infinite interactions}
In order to compute the one-body density matrix  in the infinitely repulsive case  $g\to\infty$, we start from the exact solution for the many-body wave function (\ref{eq:Psi}). Few derivation steps allow to reduce the many-body integration in (\ref{rhonu})  to the calculation of single-particle integrals: 
we label the $N!$ permutations $P\in S_N$
by an index $i_k$, where $i\in\{1,\dots,N\}$ denotes the position of the first particle, and $k\in\{1,\dots,(N-1)!\}$ labels the permutations of the $N-1$ other 
particles. Then, choosing $x\le y$, Eq.~(\ref{rhonu}) becomes
\begin{equation}
\label{eq:rhonu_ginfty}
\rho_{\nu}(x,y)=N_{\nu}\sum_{1\le i\le j\le N} R_{\nu}^{(ij)}(x,y),
\end{equation}
where
\begin{equation}
R_{\nu}^{(ij)}(x,y)=\int_{S^{(ij)}}\mathrm{d}x_2\ldots\mathrm{d}x_N\Psi(x,x_2,\ldots,x_N)\Psi(y,x_2,\ldots,x_N),
\end{equation}
with $S^{(ij)}=\{x_2<\dots<x_{i-1}<x<x_{i+1}<\ldots<x_{j-1}<y<x_{j+1}<\ldots<x_N\}$. Noticing that 
$\int\mathrm{d}x_2\ldots\mathrm{d}x_N\theta_{P_{i_k}}(x,\ldots,x_N)\theta_{P_{i_l}}(y,\ldots,x_N)\propto\delta_{kl}$, we obtain after some algebra 
\cite{Forrester03,Decamp2016}:
\begin{equation}
R_{\nu}^{(ij)}(x,y)= C_{ij}\sum_{P,Q\in S_{N-1}}\epsilon(P)\epsilon(Q)\displaystyle{\prod_{l=2}^{N} }  \int_{L_{ij}}^{U_{ij}} \mathrm{d}x_l    
(x_l-x)(x_l-y) \phi_{P(l)-1}(x_l) \phi_{Q(l)-1}(x_l),
\end{equation}
where $C_{ij}=\frac{2^{N-1}}{\sqrt{\pi}N!(N-1)!}\cdot\frac{\sum_{k=1}^{(N-1)!}a_{i_k}a_{j_k}}{(i-1)!(j-i)!(N-j)!}$,
and where the integration limits $(L_{ij},U_{ij})$ are given by
\begin{equation}
(L_{ij},U_{ij})=\left\{
\begin{array}{lll}
(-\infty,x) & \mbox{if } l\le i, \\ (x,y) & \mbox{if  } i<l\le j, \\ (y,+\infty) & \mbox{if  } l>j.
\end{array}\right.
\end{equation}

The density profile is readily obtained as 
$n_\nu(x)=N_{\nu}\sum_i R_{\nu}^{(ii)}(x,x)$, and the momentum distribution is obtained by numerically evaluating the Fourier transform (\ref{eq:defnk}) for the specific case of  Eq.(\ref{eq:rhonu_ginfty}).

The exact solution allows also for an independent evaluation of the Tan's contact using the Tan relation on the interaction energy.
Using  Eq.~(\ref{eq:K}), we find that the interaction energy  $H_{\mathrm{int},\nu\nu^\prime}$ is given by
\begin{equation}
\label{exacont}
H_{\mathrm{int},\nu\nu^\prime}=\left\{
\begin{array}{ll}
 \frac{\hbar^4}{2m^2 g}\sum_{k=1}^{N-1}\sum_{P\in \sigma_N(\nu,\nu',k)}(a_P-a_{\tau_k\circ P})^2\alpha_k & \mbox{if } \nu\ne\nu^\prime, \\ 
 \frac{\hbar^4}{2m^2 g}\sum_{k=1}^{N-1}\sum_{1\le i<j \le N_{\nu}}\sum_{P\in \sigma_N(i,j,k)}(a_P-a_{\tau_k\circ P})^2\alpha_k & \mbox{if  } \nu = \nu^\prime,
\end{array}\right.
\end{equation}
where $\sigma_N(\mu,\nu,k)$ is the set of all permutations such that particles in $k$ and $k+1$ positions are from the species $\mu$ and $\nu$
(or are particles $i$ and $j$ for $\sigma_N(i,j,k)$), 
and $\tau_k$ is the transposition of these two particles. Note that $H_{\mathrm{int},\nu\nu}=0$ if species $\nu$ is fermionic.
The calculation of the interaction energy allows to obtain the Tan's contact using Eq.(\ref{eq:tanrel}).

\subsection{Numerical solution at arbitrary interactions}

In realistic  experimental conditions, the interactions are tunable, allowing to explore all the interactions regime from small to large values. 
In particular, a relevant question is to what extent the density profiles and the momentum distributions evaluated from the exact solution at $g\rightarrow\infty$ describe accurately the ground states at large but finite interaction.
In order to tackle the arbitrary interaction regime,  we have performed numerical calculations using a two-site optimization scheme, according to the Density-Matrix Renormalization-Group (DMRG) method, based on matrix product states (MPS). The code allows the exploitation of Abelian symmetries (here particle-number conservation for each species). To address the continuous problem, we discretized the relevant region of the trap and calculated properties of the resulting lattice model. The continuum results were then obtained by increasing the number of lattice points in consecutive simulations (thereby decreasing the lattice spacing) and finite-size scaling the quantities of interest. Density distributions were measured as site-occupations in the discretized model; the Tan contacts were determined through the interaction part of the energy. The momentum distributions were obtained from Fourier-transformed two-point correlators $n_\nu(k)=\frac{1}{L}\sum_{j,l=1}^L e^{i(j-l)k}\langle c_{\nu,j}^\dagger c_{\nu,l}\rangle$, where $c_{\nu,j}^{(\dagger)}$ are the annihilation (creation) operators for the species $\nu$ in the lattice model. The maximal number of lattice sites considered was 216, covering a region of 12 harmonical trap lengths. The virtual bond-dimension of the MPS was $m\approx 150$ which appears of be sufficient for the very low density of particles we consider here.

%%%%%%%%%%%%%%%%%%%%%%%%%%%%%%%%%%%%%%%%%%%%%%%%%%%%%%%%%

\subsection{Density profiles}
The density profiles of a multi-component mixture display a wealth of information concerning the intercomponent interactions as well as the magnetic structure.

\begin{figure}
  \includegraphics[width=0.33\linewidth]{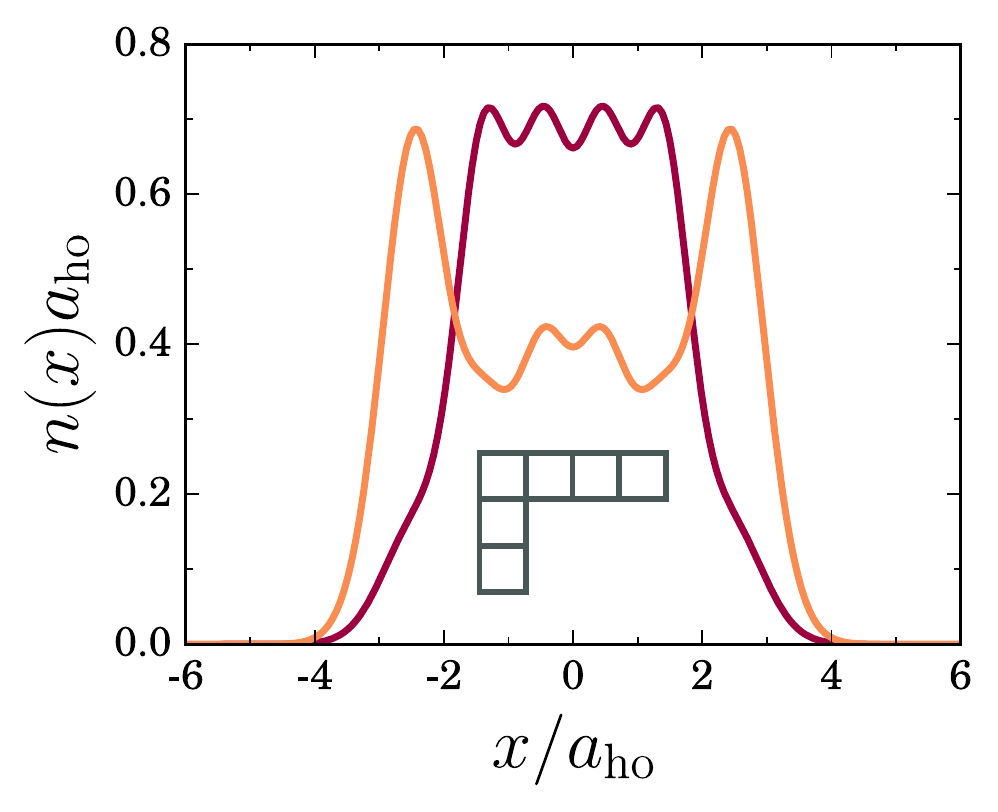}  
  \includegraphics[width=0.33\linewidth]{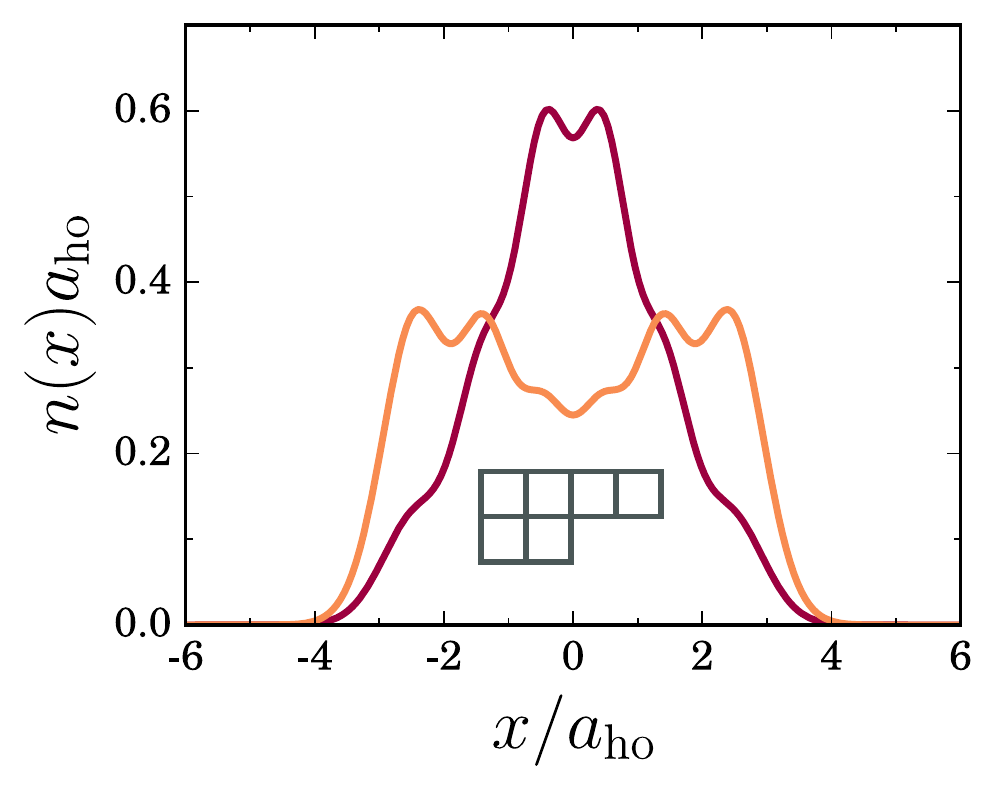}
  \includegraphics[width=0.33\linewidth]{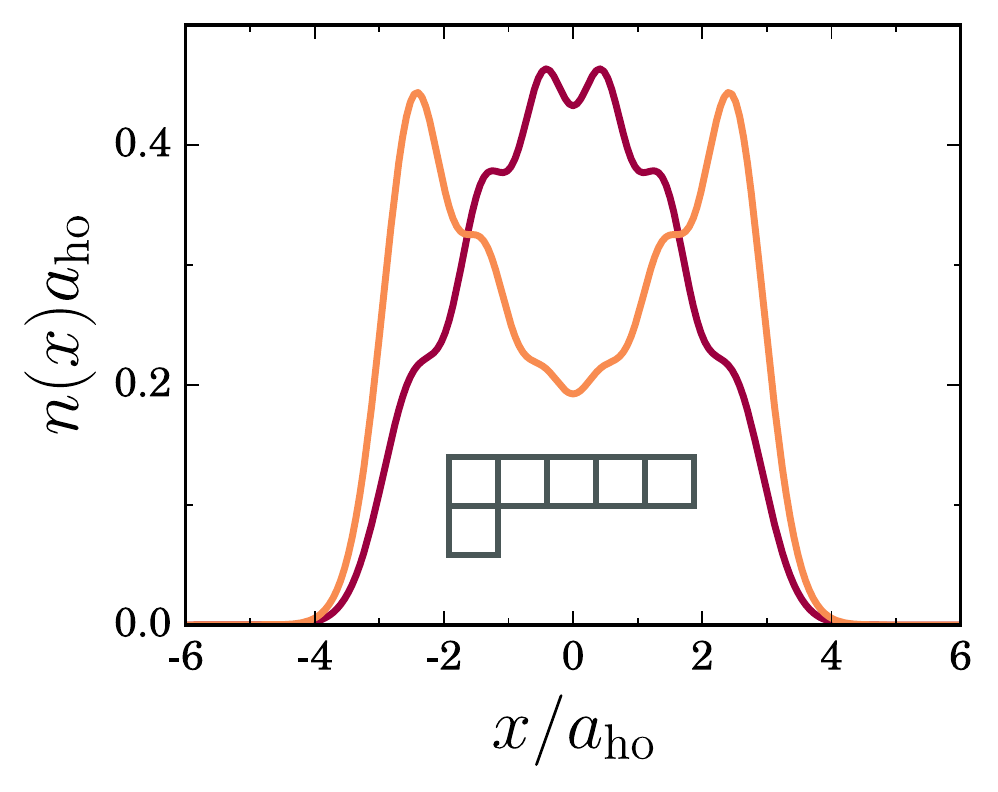}
  \caption{(Color online.) Exact solution for the bosonic (brown) and fermionic (orange) ground state density profiles (in unit of $a_{ho}^{-1}$)  in the limit $g\to \infty$,  as functions of the spatial coordinate $x/a_{ho}$. 
  From left to right, the mixtures are $3^B+3^F$, $2^B+2^F+2^F$ and $2^B+2^B+2^F$. For the case of ternary mixtures, the density profiles of the two components with the same statistics and same particle number coincide\label{fig:dens}.}
\end{figure}

At mean-field level, by increasing the strength of repulsive intercomponent interactions \cite{Das03} the system is predicted to undergo spatial separation. However, for strong interactions mean-field theory loses its validity due to the change of equation of state and the increased effect of quantum fluctuations. Luttinger liquid theory for a homogeneous system  also predicts an instability towards spatial separation  \cite{CazHo03}. For a binary mixture in a harmonic trap,  an analysis based on the local-density approximation of the exact equation of state was performed in \cite{ImaDemAnn,ImaDem06}, yielding a partial spatial separation of the two components at strong interactions. For small particle numbers, and finite interactions, spatial separation was found within some parameter regimes using mapping to a spin model and exact diagonalization in \cite{Deuretzbacher2017}, and for finite interactions  an exact solution showing spatial separation was obtained in \cite{Zinner2017}.

\begin{figure}
  \includegraphics[width=0.33\linewidth]{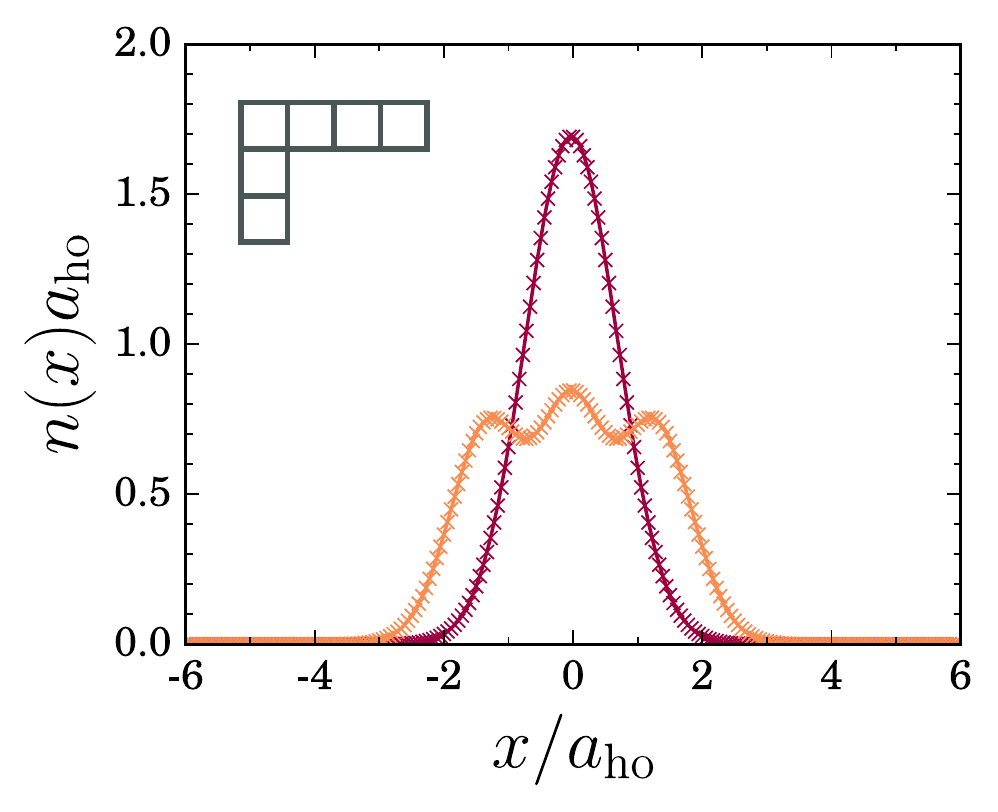}  
  \includegraphics[width=0.33\linewidth]{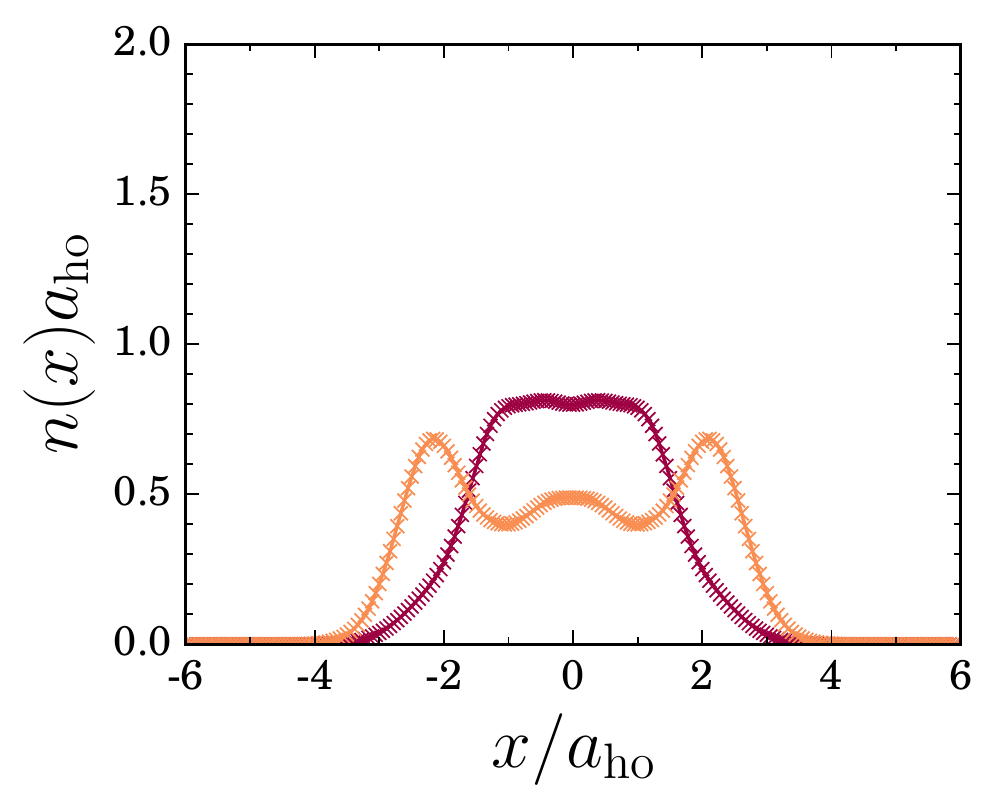}
  \includegraphics[width=0.33\linewidth]{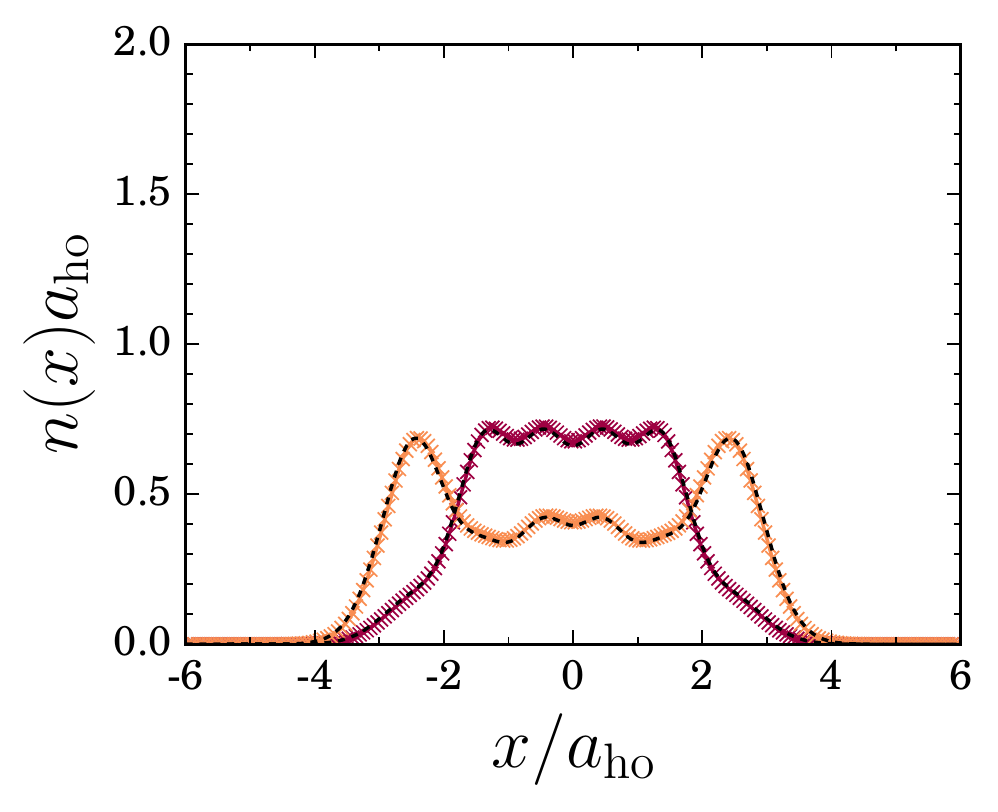}\\
  \includegraphics[width=0.33\linewidth]{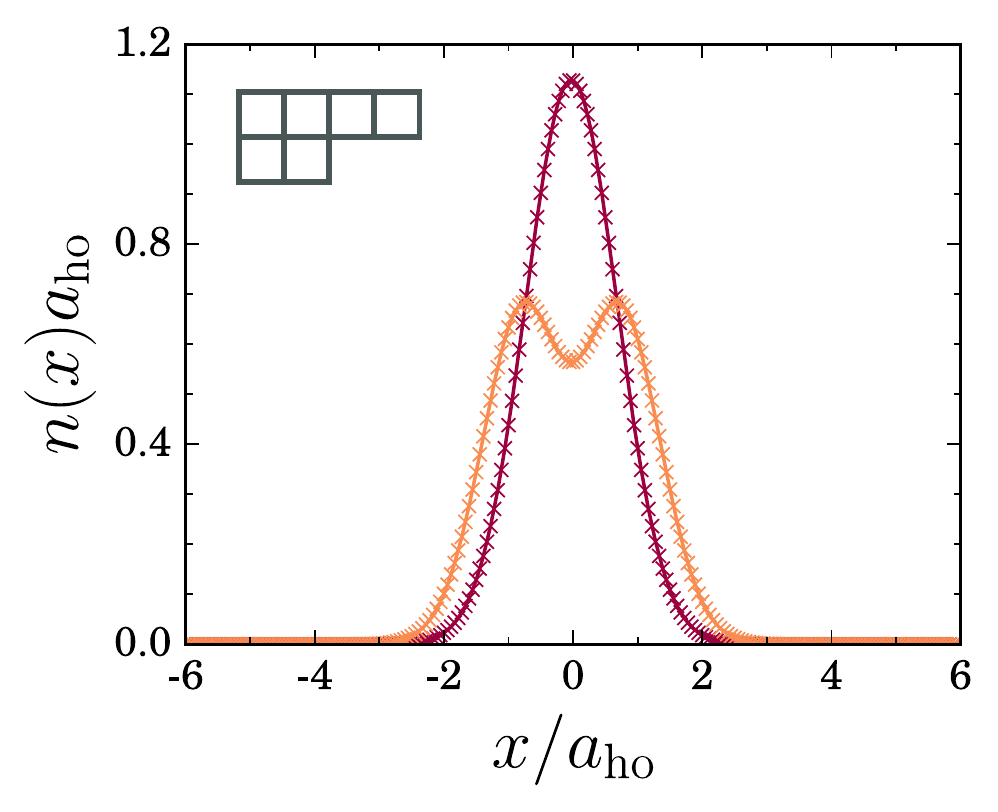}  
  \includegraphics[width=0.33\linewidth]{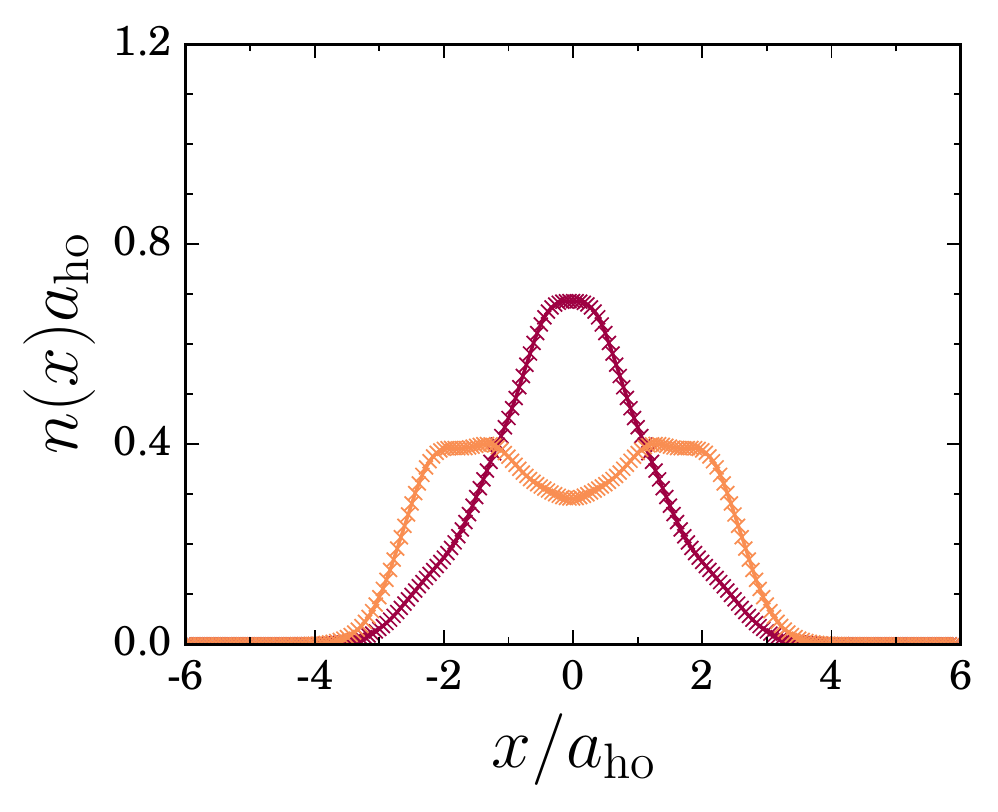}
  \includegraphics[width=0.33\linewidth]{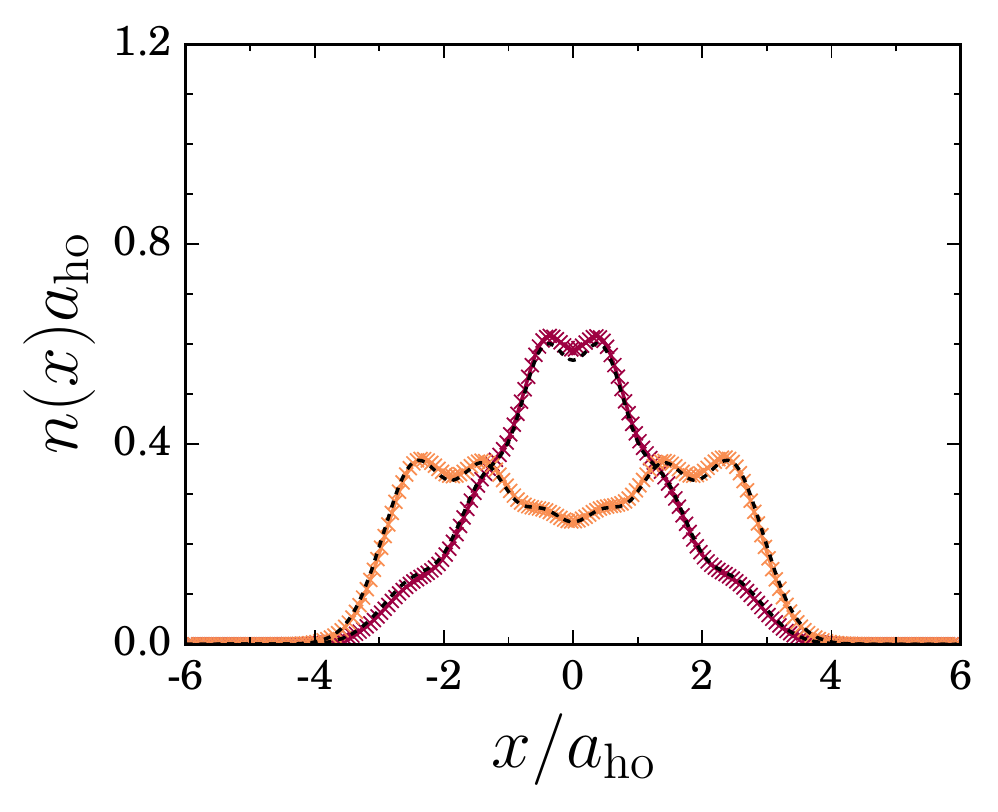}
  \caption{(Color online.) Bosonic (brown) and fermionic (orange) ground state density profiles of a $3^B+3^F$ mixture (top row) and  $2^B+2^F+2^F$ mixture (bottom row) with interactions $g=0.0\,\hbar\omega a_{ho}$ (left), $g=10.0\,\hbar\omega a_{ho}$ (center), and $g=100.0\,\hbar\omega a_{ho}$ (right). In the right panel, the exact solution for $g\rightarrow\infty$ in also shown in dashes. }
\label{fig:dens_finiteg}
\end{figure}

We present  in Figs.~\ref{fig:dens}-\ref{fig:dens_finiteg} our results for both binary and ternary multi-component mixtures made of six particles. The results for the binary mixtures agree with \cite{Zinner2017,Deuretzbacher2017} in showing spatial separation, with the bosonic component occupying the center of the trap and the fermionic component pushed towards the edges.
\footnote{The results obtained from DMRG in \cite{Fang2011} for a binary mixture at $g=1000\,\hbar\omega a_{ho}$ differ from the above, as an excited state has been caught by the numerics because of the almost degeneracy at very large interactions.}

The results on spatial  separation can be seen as a direct consequence of the generalized Lieb-Mattis theorem. Indeed, if the ground state is the most symmetric configuration allowed  by the type of mixture, then it is built in such a way to minimize the number of anti-symmetric exchanges, which is the case if the fermions are on the edges.

For ternary mixtures, at fixed numbers of particles in each component, we find that the details of the density profiles depend on the type of mixture. This shows that the spatial separation  is not only  an effect  of the interactions but also of the symmetry.  Also, we remark that the results at finite interactions display the same features as the strongly-correlated regime already at intermediate interactions. This suggests that 
the spatial symmetry is already present for finite  interactions.

\subsection{Momentum distributions and Tan's contact}

\begin{figure}
  \includegraphics[width=0.33\linewidth]{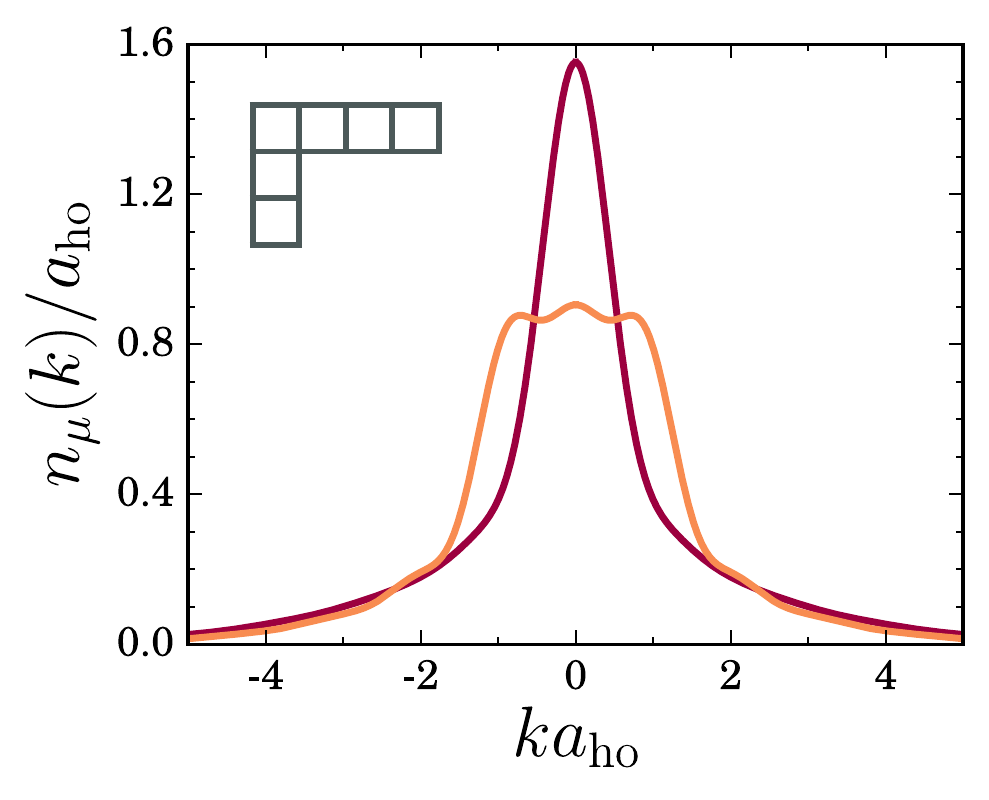}  
  \includegraphics[width=0.33\linewidth]{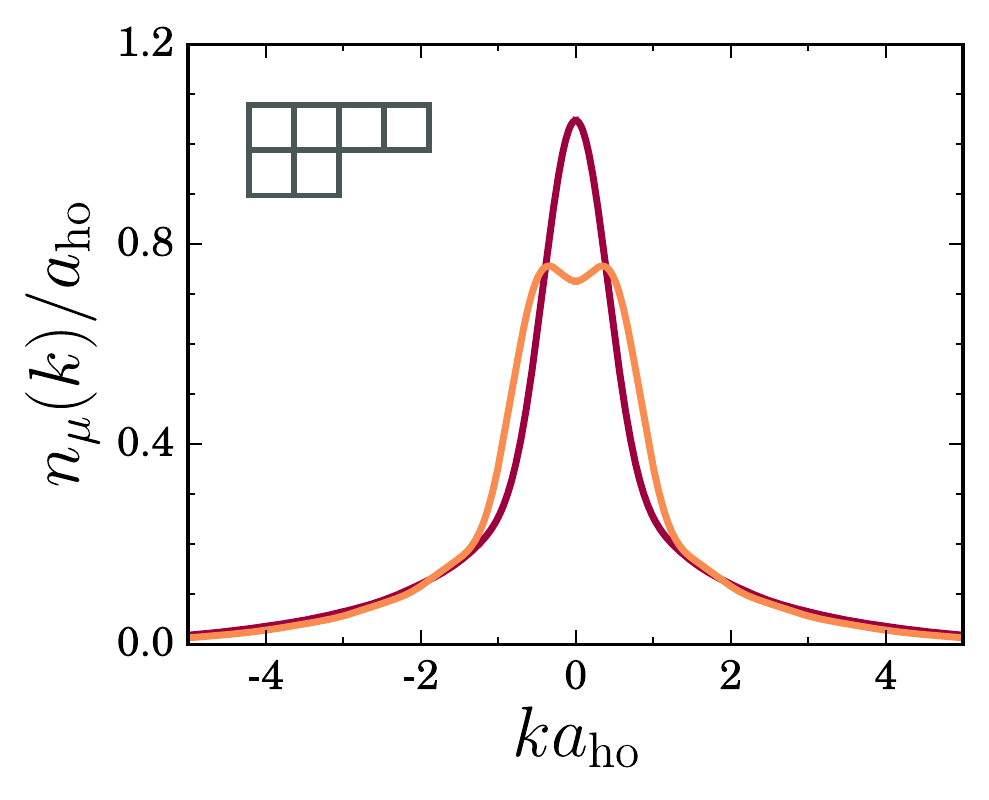}
  \includegraphics[width=0.33\linewidth]{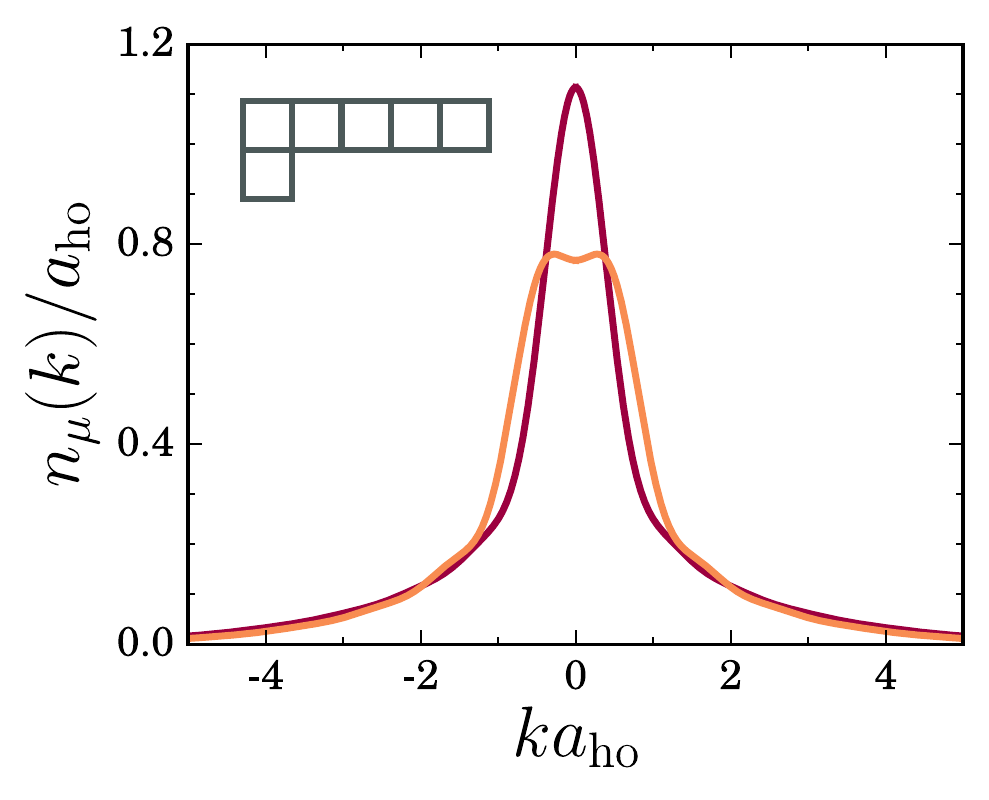}
  \caption{(Color online) Exact solution for bosonic (brown) and fermionic (orange) ground-state momentum distributions (in unit of $a_{ho}$) and as functions of $ka_{ho}$ for $g\to\infty$.
  From left to right, the mixtures are $3^B+3^F$, $2^B+2^F+2^F$ and $2^B+2^B+2^F$.\label{fig:mom}}
\end{figure} 

\begin{figure}
  \includegraphics[width=0.24\linewidth]{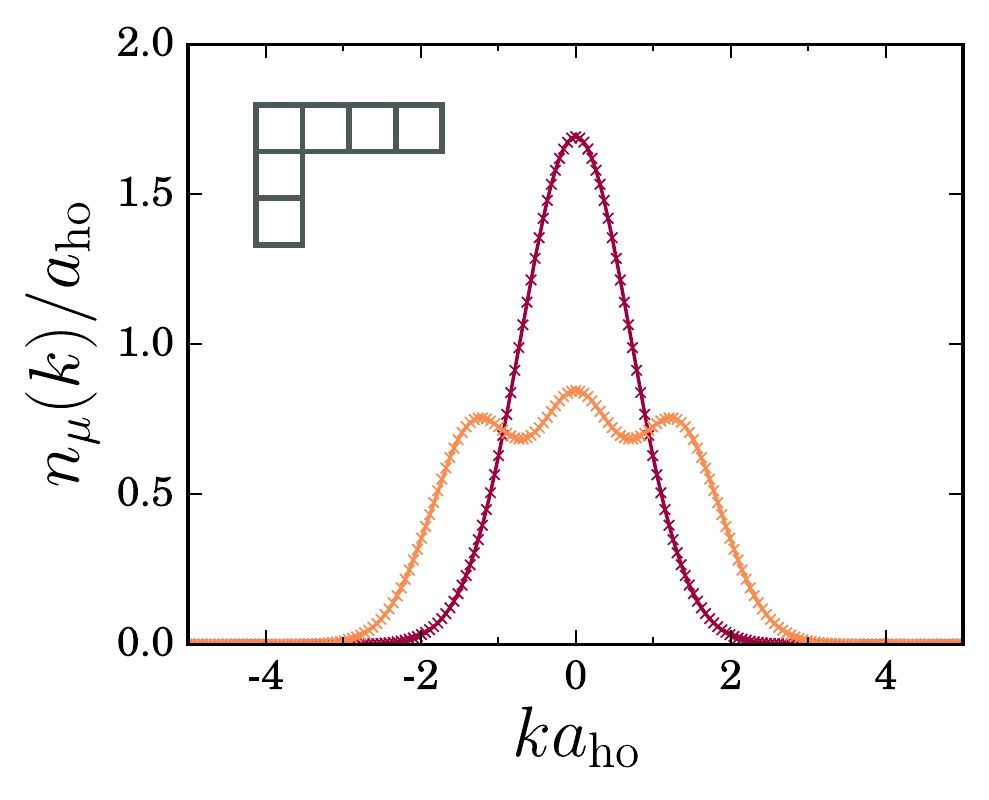}  
  \includegraphics[width=0.24\linewidth]{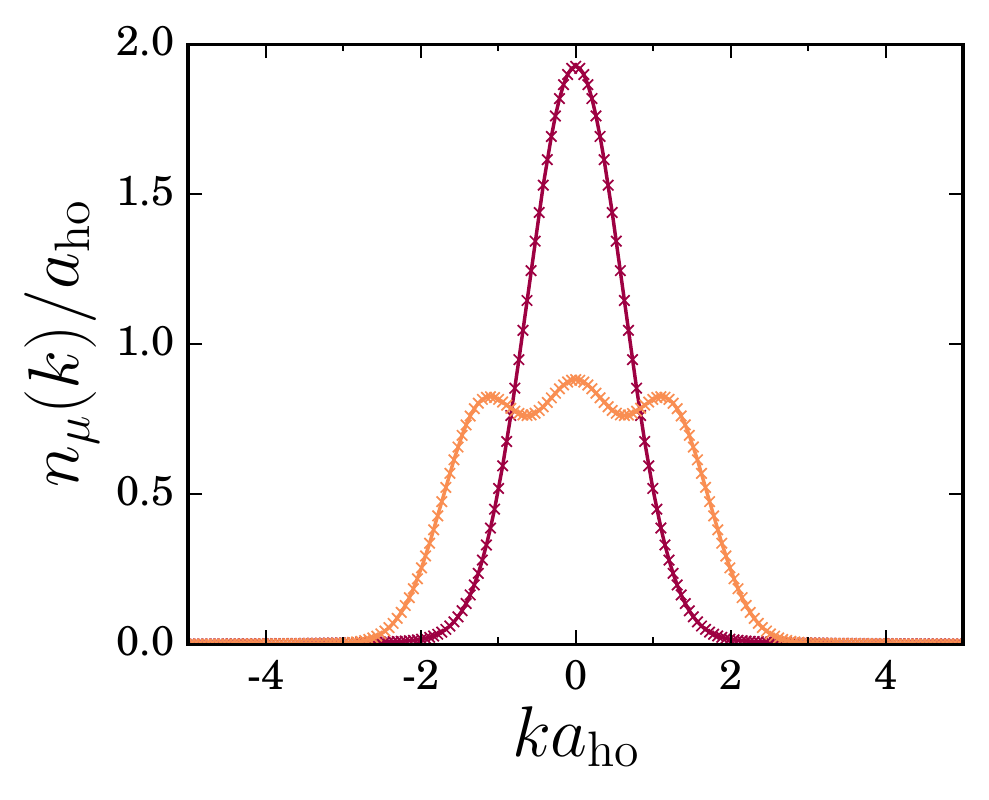}  
  \includegraphics[width=0.24\linewidth]{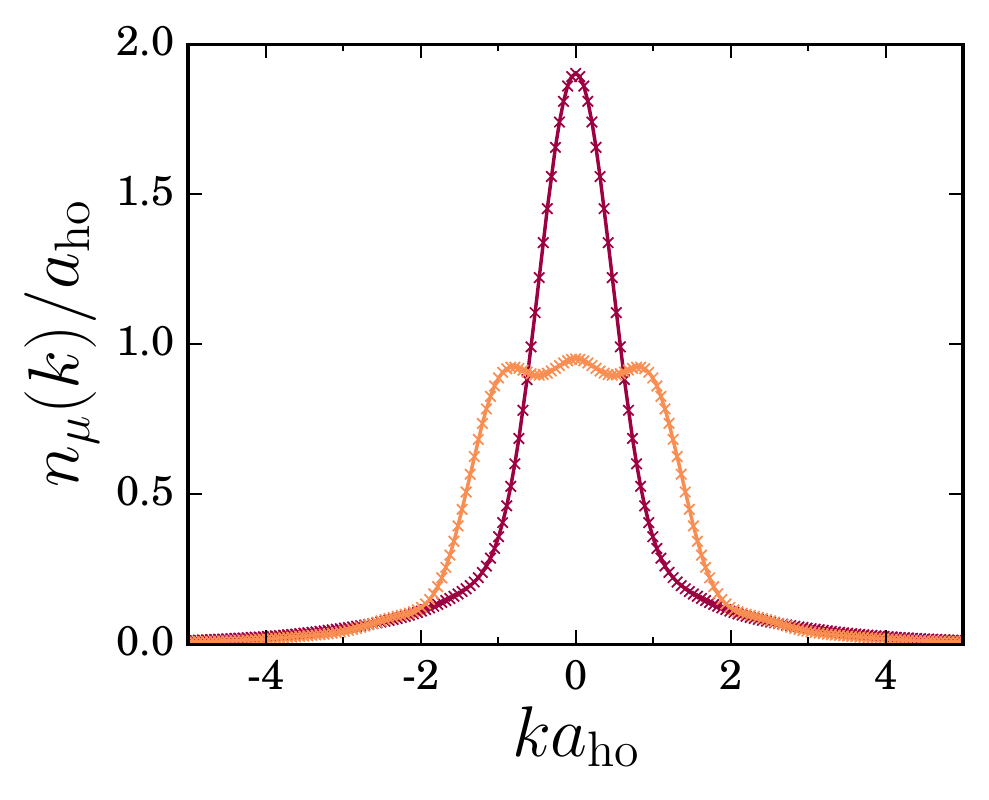}
  \includegraphics[width=0.24\linewidth]{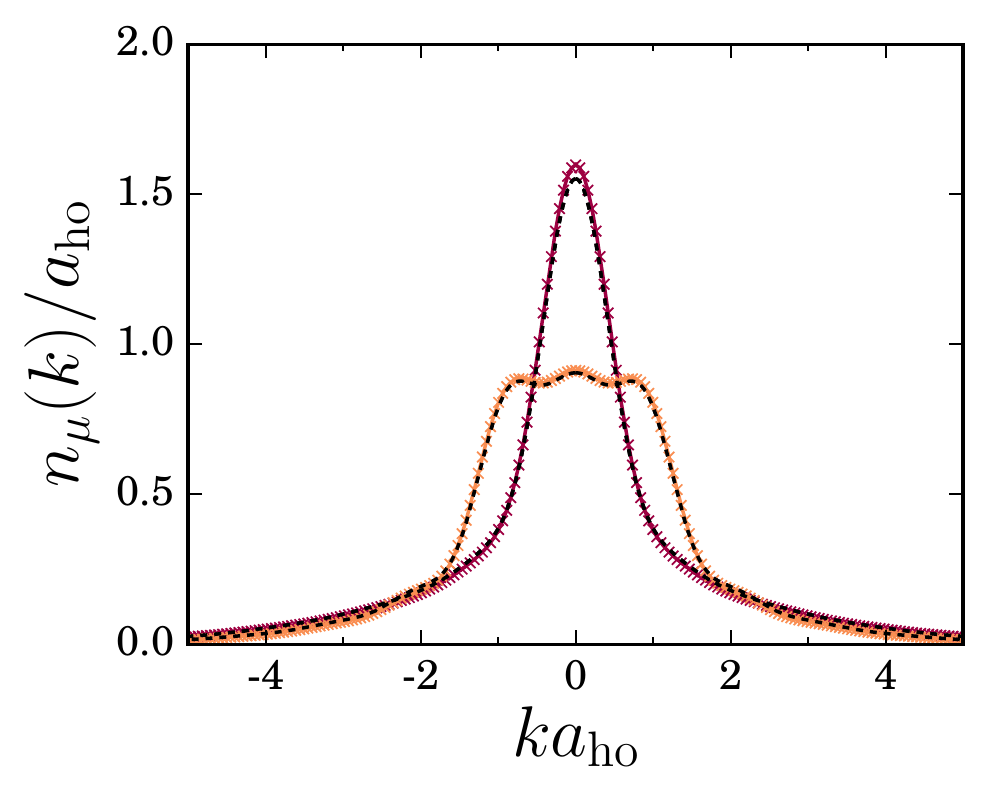}\\
  \includegraphics[width=0.24\linewidth]{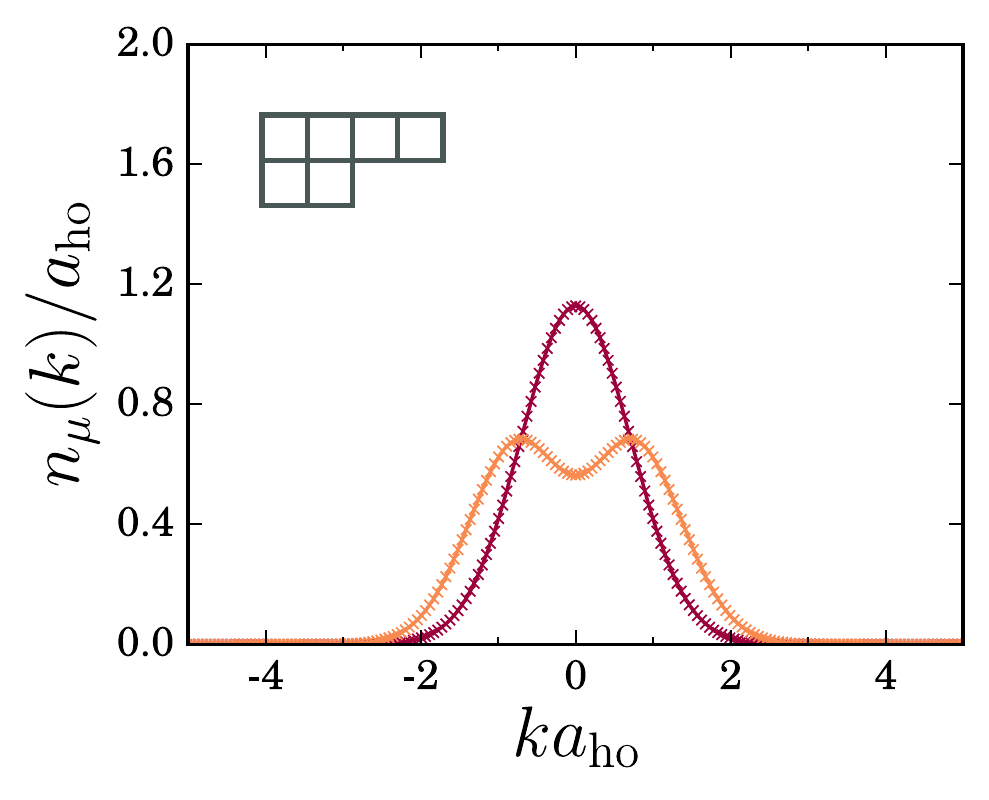}  
  \includegraphics[width=0.24\linewidth]{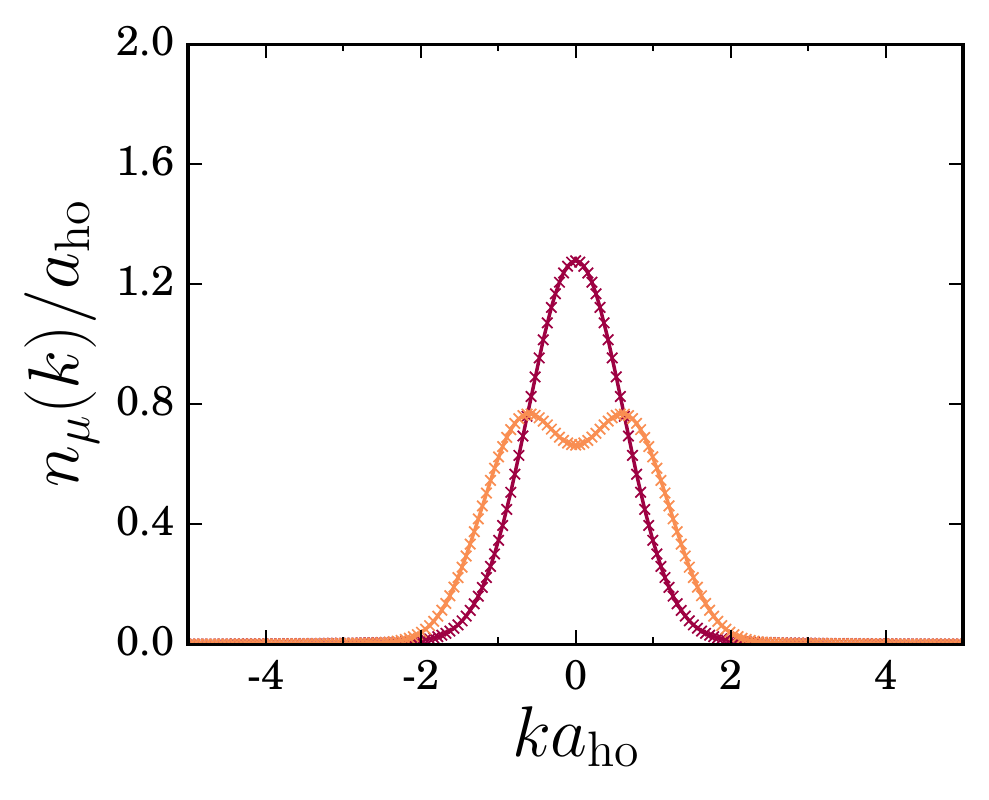}  
  \includegraphics[width=0.24\linewidth]{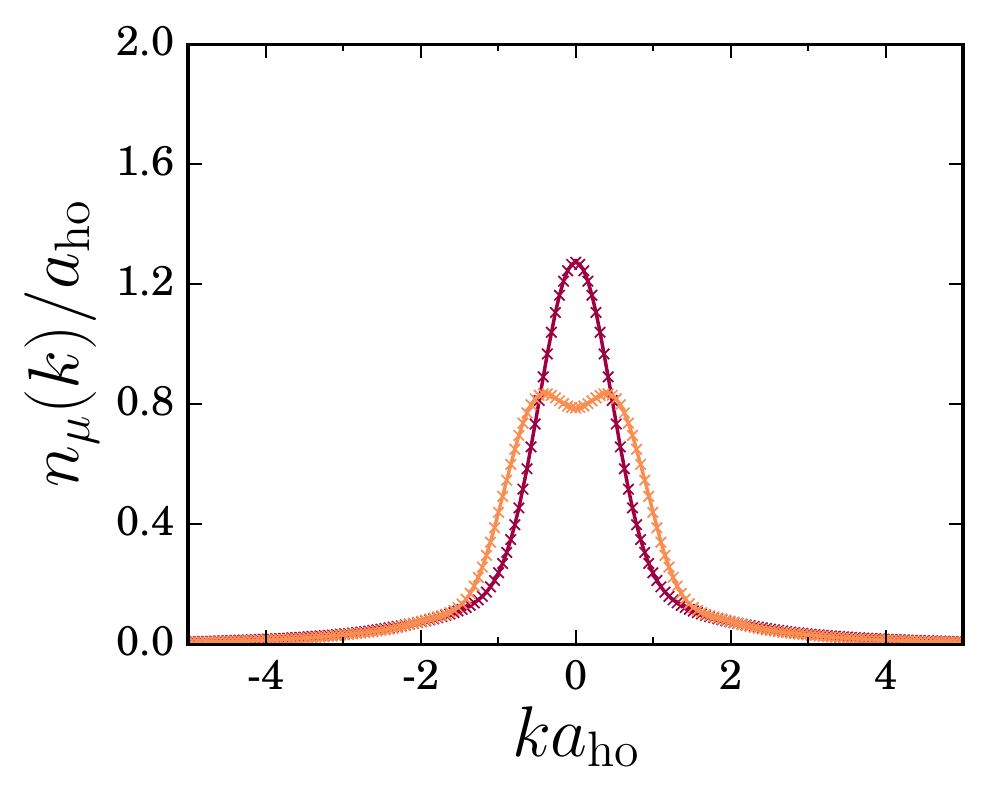}
  \includegraphics[width=0.24\linewidth]{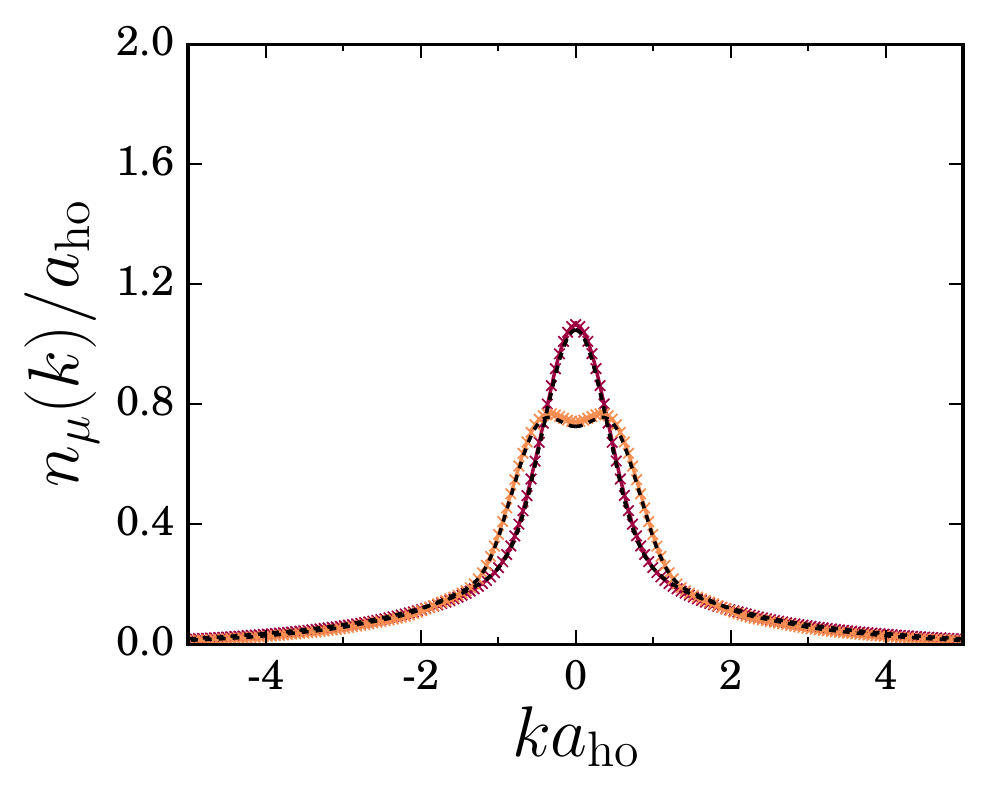}
  \caption{(Color online) Bosonic (brown) and fermionic (orange) ground state momentum profiles of a $3^B+3^F$ mixture (top row) and a  $2^B+2^F+2^F$ mixture (bottom row) with interactions $g=0.0\,\hbar\omega a_{ho}$, $g=1.0\,\hbar\omega a_{ho}$, $g=10.0\,\hbar\omega a_{ho}$ and $g=100.0\,\hbar\omega a_{ho}$ (from left to right). In the right panel, the exact solution for $g\rightarrow\infty$ in also shown in dashes.}
\label{fig:mom_finiteg}
\end{figure}

We now turn to the discussion of our results for the momentum distribution. Fig.~\ref{fig:mom} shows our predictions for the regime of infinite interaction strength. The momentum distribution of the bosonic components display one central peak, whereas the fermionic components display as many peaks as the number of the fermions in the corresponding species. This feature, typical of non-interacting spinless fermions in harmonic confinement \cite{Vignolo00}, has also been reported for  multi-component interacting fermionic mixtures in harmonic traps  \cite{Deuretzbacher2016}. The occurrence of different peaks  in the fermionic momentum distribution can be seen as a  consequence of the Pauli principle for fermions belonging to the same species.

\begin{figure}
\begin{center}
 \includegraphics[width=0.33\linewidth]{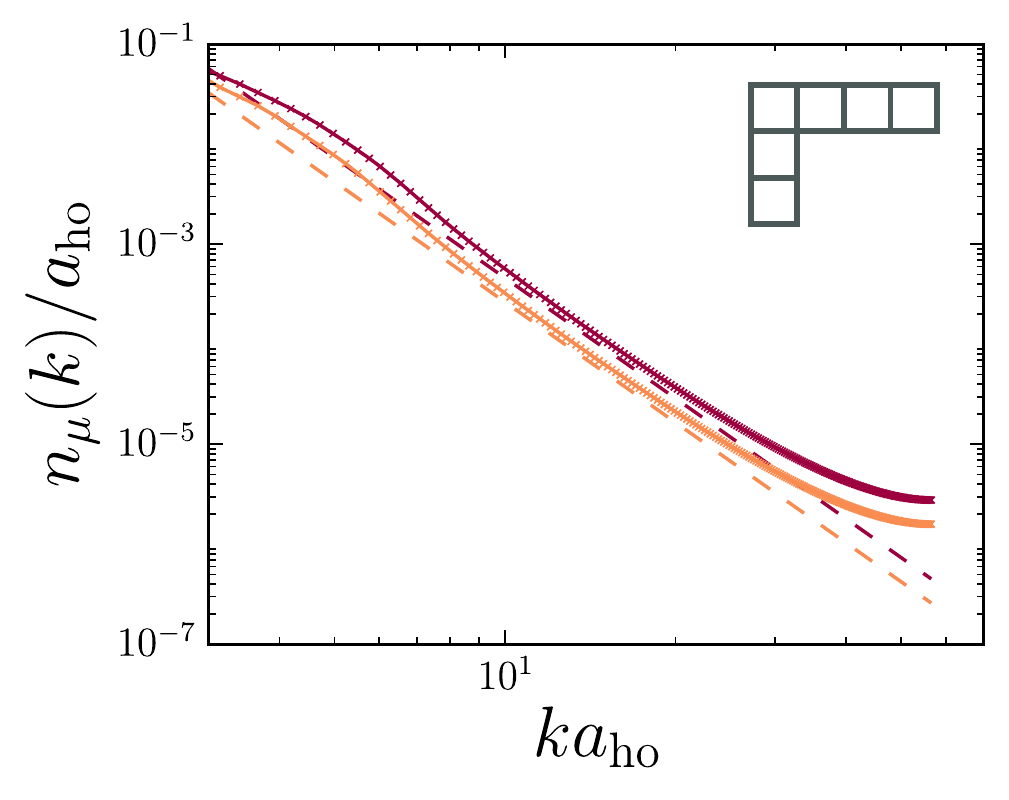}
 \includegraphics[width=0.33\linewidth]{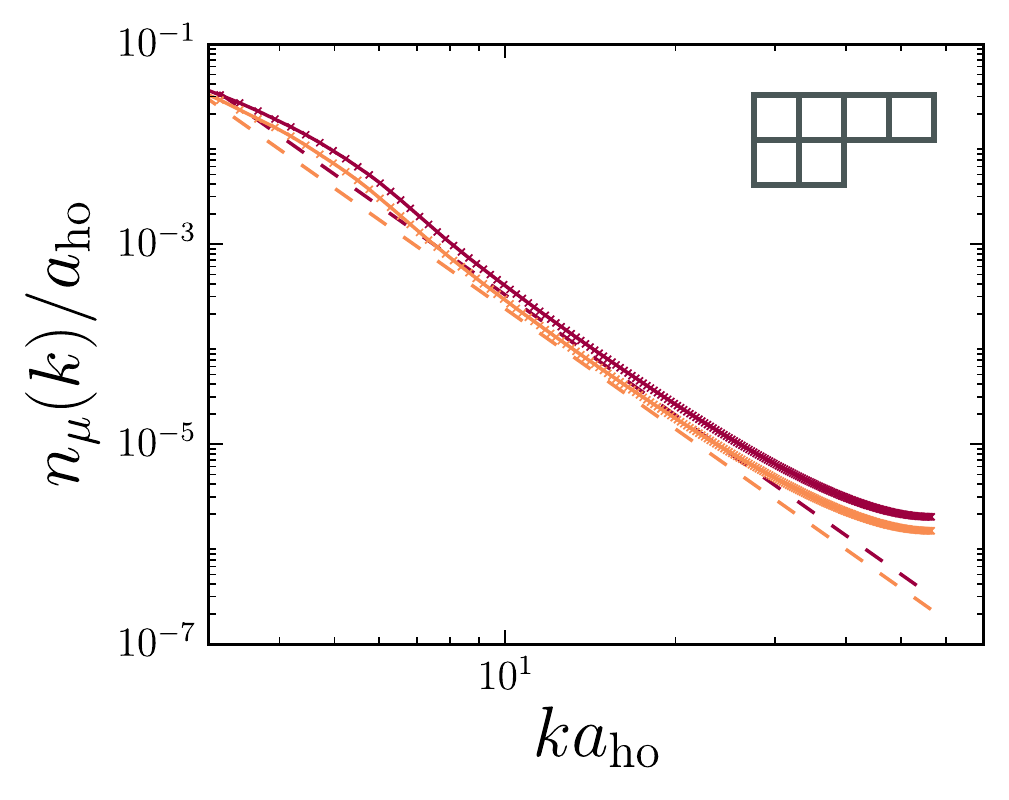}
 \end{center}
  \caption{(Color online) Tails of the momentum distributions (brown: bosons, orange: fermions) in a $3^B+3^F$ mixture (left) and in a  $2^B+2^F+2^F$ mixture (right), with interactions $g=10.0\,\hbar\omega a_{ho}$. The dashed lines show the Tan contacts as determined through the interaction energies in Eq.~\eqref{eq:tanrel}. The bending of the momentum curves for very large momenta is an artifact originating from the discretization procedure.}
\label{fig:contact_finiteg}
\end{figure}

\begin{figure}
  \includegraphics[width=0.33\linewidth]{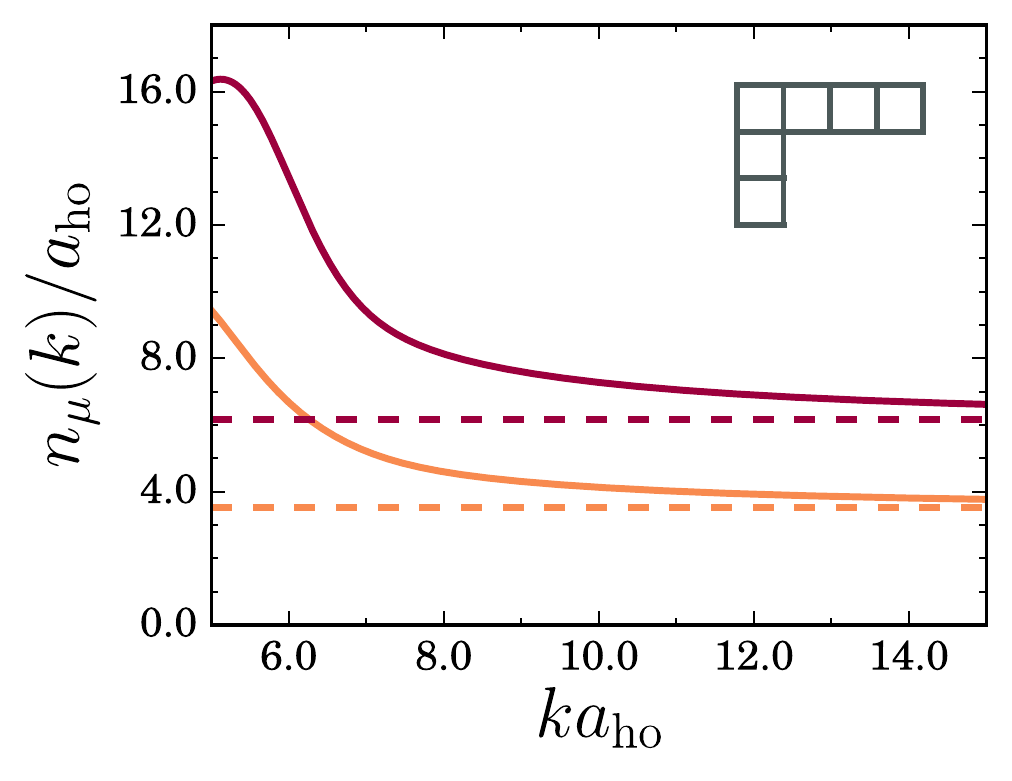}  
  \includegraphics[width=0.33\linewidth]{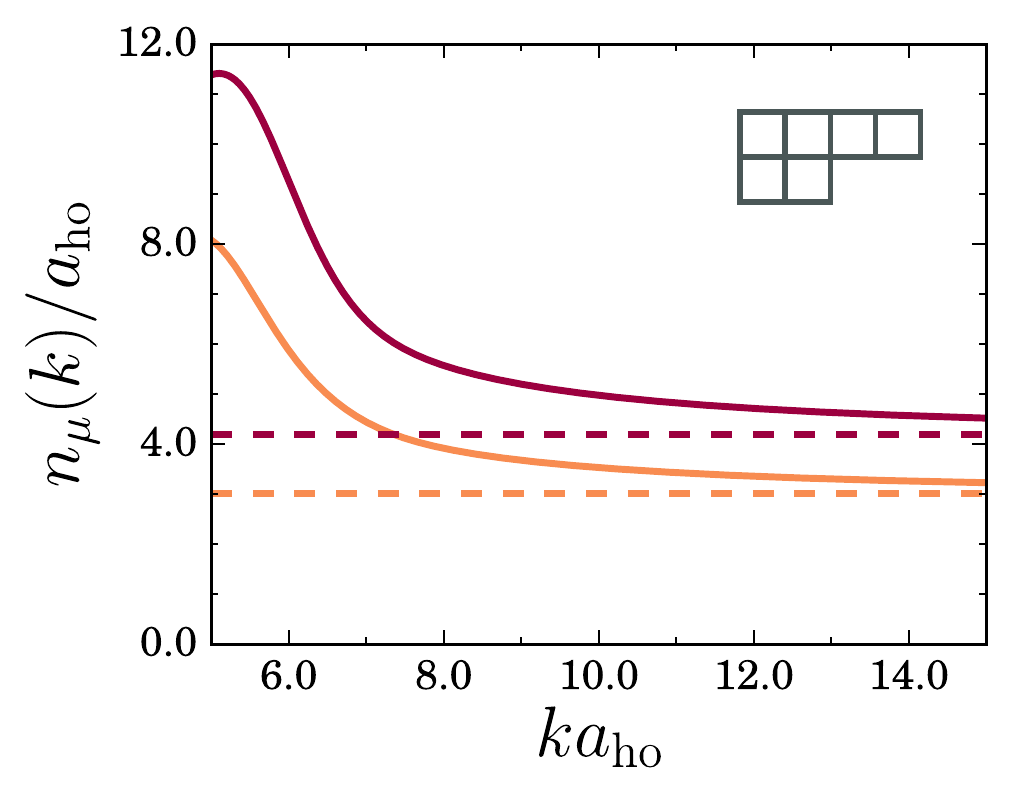}
  \includegraphics[width=0.33\linewidth]{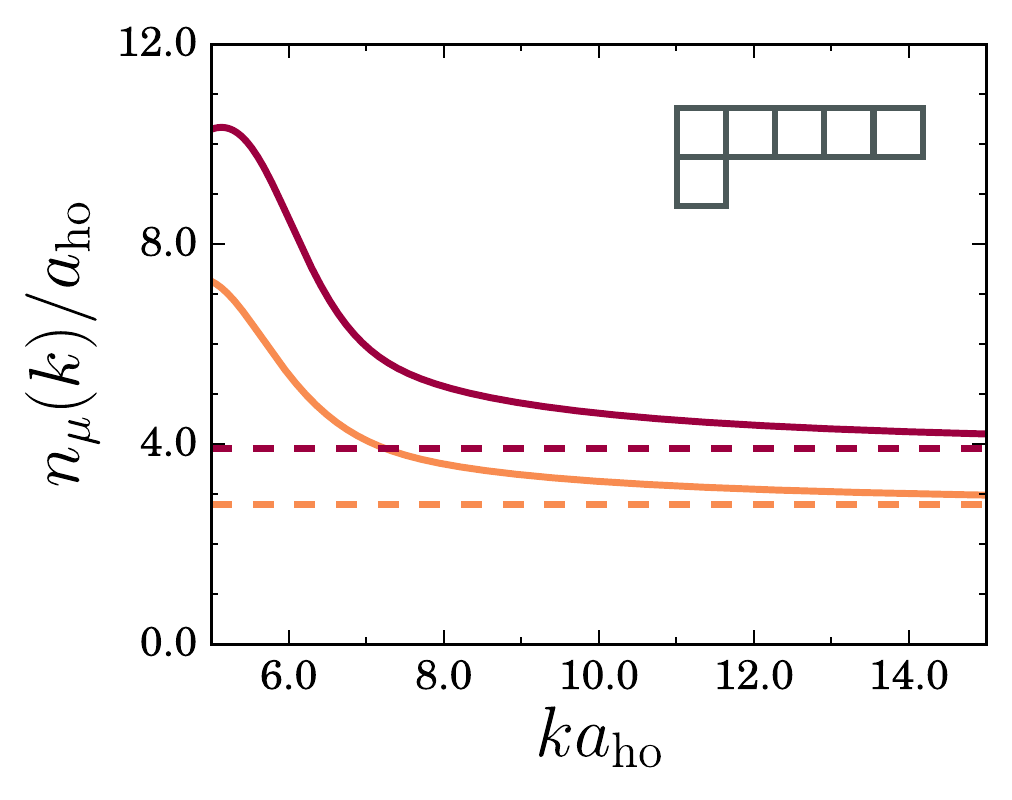}
  \caption{(Color online) Bosonic (brown) and fermionic (orange) ground state $k^4n_{\nu}(k)$ functions in unit of $a_{ho}^{-3}$ and as functions of $ka_{ho}$. 
  From left to right, the mixtures are $3^B+3^F$, $2^B+2^F+2^F$ and $2^B+2^B+2^F$.
  The horizontal lines are obtained from the exact solution via Eq.~(\ref{exacont}) in Eq.(\ref{eq:tanrel}). \label{fig:contact}}
\end{figure}

Fig.~\ref{fig:mom_finiteg} shows the dependence of the momentum distributions on interaction strength. At increasing interactions we notice that the number of peaks in each distribution is conserved, and the shape of both fermionic and bosonic momentum distributions looks qualitatively the same. However, a detailed analysis shows that both bosonic and fermionic momentum distributions shrink in their central part. This is complementary to the coordinate space description, where repulsive interactions tend to broaden the profiles. Furthermore, the tails of the momentum distribution become more and more important at increasing interactions, as it is visible on Fig.~\ref{fig:mom_finiteg}. These tails are shown in Fig.~\ref{fig:contact_finiteg} for a finite value of the interaction parameter $g=10.0\,\hbar\omega a_{ho}$ and in Fig.~\ref{fig:contact} for the limiting case $g\rightarrow\infty$. All the mixtures analysed exhibit the $k^{-4}$ behavior with the expected Tan's contact coefficients. The asymptotic behaviour of the tails (dashed lines) has been obtained from Eq.(\ref{eq:tanrel}), where the interaction energy for the mixtures at $g=10.0\,\hbar\omega a_{ho}$ has been numerically calculated (Fig.~\ref{fig:contact_finiteg}), while for the mixtures at $g\rightarrow\infty$ has been evaluated from the exact solution via Eq.~(\ref{exacont}) (Fig.~\ref{fig:contact}). For balanced mixtures with the same number of particles in each component we notice that Tan's contact are not equal for bosonic and fermionic species, at difference from the case of multi-component fermionic mixtures \cite{Decamp2016-2}. This properties readily follows from Eq.~(\ref{exacont}), showing that in the bosonic case there is an additional term corresponding to intracomponent boson-boson interactions.

Finally, we notice that the knowledge  of the Tan's contacts $\mathcal{C}_{\nu}$ for each component of the mixture allows to deduce the symmetry of the mixture, since   the total contact 
$\mathcal{C}_{\mathrm{tot}}=\sum_{\nu=1}^{\mathrm{b+f}}\mathcal{C}_{\nu}$ 
is linked to the energy slope $K$ via Eq. (\ref{eq:K}) and (\ref{eq:tanrel}) through
\begin{equation}
\mathcal{C}_{\mathrm{tot}}=\frac{m^2}{\pi\hbar^4}K,
\end{equation}
and each $K(S)$ corresponds to a unique symmetry $S$ (see Table \ref{tab:sym}  in Sec.~\ref{sec:Symmetry}). Thus, an experimental measurement of Tan's contact, e.g.\@ through
spin-resolved time-of-flight measurements, or through the energy, would allow a symmetry spectroscopy of the Bose-Fermi mixture \cite{Wild2012}. 
This generalizes our previous results on fermionic mixtures \cite{Decamp2016-2}.

%%%%%%%%%%%%%%%%%%%%%%%%%%%%%%%%%%%%%%%%%%%%%%%%%%%%%%%%%

\section{Conclusions}
\label{sec:Concl}
In this work we have studied the spatial and the momentum distributions of a zero-temperature multi-component boson-fermion mixture, trapped in a one-dimensional harmonic potential.
We have considered the ideal case of equal-mass particles and equal contact-interaction strength, that could be obtained in a  good approximation in the experiments by trapping and cooling large atomic-number isotopes (e.g.\@ Ytterbium). At infinitely large repulsive interactions we have determined the ground-state properties by  using an exact many-body wavefunction. 
At finite interactions, we have calculated the spatial density profiles and momentum distributions by numerical DMRG methods, obtaining an excellent agreement between the exact calculations  and the numerical ones. We have found that at large interactions  bosons and fermions are spatially phase separated in the trap and that the momentum distribution tails - the contacts - at fixed number of particles, increase by increasing the number of bosons and/or the number of fermionic components. Both effects are related to the wavefunction symmetry, that we have shown to be, by exploiting a sum-class method, the most possible symmetric one for each type of mixture.
Our results are in agreement with a generalized version of the Lieb-Mattis theorem. Importantly, we have found that the exchange symmetry of the many-body wave-function is preserved for large but finite interactions, as in typical experimental conditions, and we have shown that this symmetry is measurable through the tails of the momentum distribution.

\section*{Acknowledgments}
We acknowledge funding from the ANR project SuperRing (ANR-15-CE30-0012-02). J.J. thanks Studienstiftung des deutschen Volkes for financial support. The DMRG simulations were run by J.J. and M.R. on the Mogon cluster of the JGU (made available by the CSM and AHRP), with a code based on a flexible Abelian Symmetric Tensor Networks Library, developed in collaboration with the group of S. Montangero at the University of Ulm.

\bibliographystyle{unsrtnat}

%\bibliography{biblioferm}

\end{document}